\documentclass[submission,copyright,creativecommons]{eptcs}
\providecommand{\event}{TARK 2019} % Name of the event you are submitting to
\usepackage{breakurl}             % Not needed if you use pdflatex only.
\usepackage{underscore}           % Only needed if you use pdflatex.

\usepackage{wrapfig}
\usepackage[export]{adjustbox}
 \usepackage{rotating}
 \usepackage{proof}
 \usepackage{url}
 \usepackage{xcolor}
 \usepackage{ifthen}
 \usepackage[normalem]{ulem}
 \usepackage{microtype} 
 \usepackage{caption}
 \usepackage{amsthm}

\usepackage{enumitem}

 \usepackage{amssymb,amsmath}
 \usepackage[all]{xy}  
 \usepackage{enumitem}
 \usepackage{bbm}
 \usepackage{amsfonts}
 \usepackage{mathtools}
 \usepackage{etex}

 \usepackage{tcolorbox}

 \input{coalgebra.sty}

 \newcommand{\reals}{\mathbb{R}}

 \newcommand{\rem}[1]{\relax}

 \newlength{\mathfrwidth}
 \newsavebox{\mathfrbox}

    \newcommand{\argmax}{\mbox{argmax}}
    \newcommand{\argmin}{\mbox{argmin}}
        
\makeatletter
\def\maketag@@@#1{\hbox{\m@th\normalfont\normalsize#1}}
\makeatother        
        
\renewcommand{\to}{\rightarrow}

\renewcommand{\o}{\circ}

\newcommand{\PhiExt}{\Phi_{\sim}}

\newcommand{\PhiD}{\Phi^{\Dis}}

\newcommand{\MExt}{M_{\sim}}

\newcommand{\vproofspace}{\vspace{-0.25cm}}

\newcommand{\Dis}{\mu}

\DeclareMathOperator{\Prob}{Pr}
\DeclareMathOperator{\Inf}{\mathcal{F}}

\DeclareMathOperator{\MostInf}{MostForce}
\DeclareMathOperator{\LeastInf}{LeastForce}

\DeclareMathOperator{\DICT}{DICT}
\DeclareMathOperator{\PIIA}{PIIA}
\DeclareMathOperator{\Dict}{Dict}

\renewcommand{\F}{\mathcal{F}}
\newcommand{\interior}[1]{%
  {\kern0pt#1}^{\mathrm{o}}%
}

\newcommand{\LA}{\mathcal{L} ( A )}

\newcommand{\piv}{\vec{\pi}}
\newcommand{\tauv}{\vec{\tau}}
\newcommand{\sigmav}{\vec{\sigma}}
\newcommand{\emphi}{}  % take this one if emph is NOT wanted in intro

\def\usesectionnumbering{1}  % set to 1 to override, any other value to use Springer numbering

\ifthenelse{ \equal {\usesectionnumbering}{1} }
{
\newcounter{nicecntr} %[section]
\numberwithin{nicecntr}{section}

\newenvironment{lemma}[1][]{\refstepcounter{nicecntr}\par\medskip
   \noindent\textbf{Lemma~\thenicecntr\ifthenelse{ \equal {#1} {} }{}{\ (#1)} } \rmfamily\it}{\medskip}

\newenvironment{definition}[1][]{\refstepcounter{nicecntr}\par\medskip
   \noindent\textbf{Definition~\thenicecntr\ifthenelse{ \equal {#1} {} }{}{\ (#1)} } \rmfamily\it}{\medskip}

\newenvironment{proposition}[1][]{\refstepcounter{nicecntr}\par\medskip
   \noindent\textbf{Proposition~\thenicecntr\ifthenelse{ \equal {#1} {} }{}{\ (#1)} } \rmfamily\it}{\medskip}

\newenvironment{theorem}[1][]{\refstepcounter{nicecntr}\par\medskip    %%% FF: I changed 'theoremx' into 'theore\MExt
   \noindent\textbf{Theorem~\thenicecntr\ifthenelse{ \equal {#1} {} }{}{\ (#1)} } \rmfamily\it}{\medskip}
}
{}

\newcounter{nicecntrr} %[section]
\numberwithin{nicecntrr}{section}

\newenvironment{thom}[1][]{\refstepcounter{nicecntrr}\par\medskip    %%% FF: I changed 'theoremx' into 'theore\MExt
   \noindent\textbf{Theorem~{(Arrow~\cite{arrowjko}) }{}}\rmfamily\it}{\medskip}

\title{Arrow's Theorem Through a Fixpoint Argument}
\author{Frank M.\,V.\,Feys \qquad\qquad Helle Hvid Hansen
\institute{Faculty of Technology, Policy and Management \\ 
Delft University of Technology\\
 Delft, The Netherlands\\
}
\email{\; \hspace{-0.54cm} f.m.v.feys@tudelft.nl \quad\quad \; \; h.h.hansen@tudelft.nl}
}
\def\titlerunning{Arrow's Theorem Through a Fixpoint Argument}
\def\authorrunning{Frank M.\,V.\, Feys \& Helle Hvid Hansen}
\begin{document}
\maketitle

\begin{abstract}
We present a proof of  Arrow's theorem from social choice theory that uses a fixpoint argument.
Specifically, we use Banach's result on the existence of a fixpoint of a contractive map defined on a complete metric space. 
Conceptually,  our approach shows that dictatorships can be seen as
 ``stable points'' (fixpoints)
of a certain process.
%Dictatorships, the ultimate result of these processes, can be seen as ``stable elements'' (fixpoints). 
    \medskip

\end{abstract}

\section{Introduction}		\label{sec-intro}

Arrow's impossibility theorem, introduced in Kenneth Arrow's seminal monograph \emph{Social Choice and Individual Values}~\cite{arrowjko}, 
deals with the issue of 
%combining
aggregating 
 the preferences of 
 a group of individuals 
% a group's individual members 
%of a group 
%over candidates 
into a single collective preference that appropriately represents the group.

Arrow defined a social welfare function to be a function that aggregates a collection of individual preferences into a societal preference, also called social choice.
Preferences are defined as linear orders, where
candidates are ranked from top to bottom. 
%Say upfront that Arrow formulates his theorem in the setting of voting rules (instead of suddenly speaking about them here as synonyms of social welafer functions). That way it becomes clear why you are talking about preferences as linear order over candidates.
%More specifically, Arrow called a function that takes as input the 
%individual preferences of the group members, 
%each linearly ordered from top to bottom, and 
%associates to it their collective preference, the so-called
%societal outcome or \emphi{social choice}, 
% a \emphi{social welfare function}. 
He considered the following three desirable properties 
 that a social welfare function, which can be thought of as an election scheme or voting rule,  might satisfy:
\begin{itemize}
\item \textit{Pareto condition.}  If all voters rank some candidate  higher than another one, then also the social choice should do so. 
%\item \textit{Independence of irrelevant alternatives.} The societal ranking of any two candidates depends  only on the relative rankings of those candidates, and on no others.
\item \textit{Independence of irrelevant alternatives.} The societal ranking of any two candidates depends  only on their relative rankings
in the individual rankings, and on nothing else.
%\footnote{In equivalent terms: the election scheme follows the Condorcet method.}
\item \textit{Non-dictatoriality.} There is no dictator, meaning that there is no voter whose individual preference always coincides with the societal outcome.
\end{itemize}
%Arrow's impossibility theorem is the statement that even these seemingly  mild conditions of reasonableness cannot be simultaneously satisfied by any voting rule. 
Arrow's impossibility theorem states that  these seemingly  mild conditions of reasonableness cannot be simultaneously satisfied by any voting rule. 

In this paper we provide a new proof of this theorem that 
uses
Banach's fixpoint theorem, which states that a contractive map on a 
complete metric space has a unique fixpoint~\cite{banach1922operations}.
%
%Banach's theorem, which states that a contractive map on a complete metric space has a unique fixpoint.
%
%uses a fixpoint argument. 
%We present a proof of  Arrow's Theorem from social choice theory that uses a fixpoint argument.
%To be precise, we use Banach's result on the existence of a fixpoint of a contractive map defined on a complete metric space~\cite{banach1922operations}.
%Roughly speaking, our approach furthers the perspective on Arrow's impossibility result 
%that
% election mechanisms necessarily converge towards a dictatorship, and
%dictatorships can be seen as ``stable points'' (fixpoints) 
%of this process. 
Our proof method shows that dictatorships can be seen as
 ``stable points'' (fixpoints)
of a certain process.

We give a sketch of our proof. 
First we define a metric 
parametrized by a  probability distribution on profiles. 
The distance between two voting rules is the probability 
under the given distribution 
that the outcome of the election is different.
%is different for the two voting rules. 
The use of a distribution 
has the benefit that it allows for 
profiles to be considered concurrently.  
Inspired by the technical notion of \emphi{influence} from Boolean analysis~\cite{o2014analysis}, we then
 introduce
a notion that we call ``force'' 
(a related idea was already presented 
in~\cite{Feys:Thesis:2015}). 
The force of a voter on a given voting rule is the probability 
that the outcome of the election 
coincides with that voter's preference. 
We use this notion to define a map that changes 
voting rules by considering the least powerful 
voters' votes and transferring them to 
the most powerful voter. 
This map is shown to be a contraction.
The process converges towards its unique fixpoint, 
because of Banach's fixpoint theorem, 
and we show this fixpoint to be 
the set of dictators. 
One technical  point that is fundamental to our approach 
is that we must consider voting rules 
only up to equivalence via permutation of the voters, 
leading to a quotient metric space.
This requires us to see to it  that
%in extending
 the original metric extends appropriately 
to the quotient metric.
A certain group action takes care of that. 
Our technical development (metric space) is parametric in the distribution, and 
we study all notions as much as possible in generality. 
In the final step, we pick a specific distribution and 
derive Arrow's theorem from it.

%First we define a metric space of Pareto voting rules, and
%% that are Pareto. 
%inspired by the technical notion of \emphi{influence} from Boolean analysis~\cite{o2014analysis}, 
%we then aim to define a contractive mapping on it  that has 
% the set of dictators as its fixpoint. 
% But in order for that to work out technically, we first have to consider an equivalence relation on the original space, while making sure 
% that the original metric extends appropriately.

% Not sure I would say that using metric spaces and convergence is high-level. I find it rather low-level. Maybe a better contrast is: combinatoric vs analytic ?
 
 Our proof does not require us to manipulate specific profiles, like 
 most other previous proofs, but offers an analytic perspective 
 in terms of fixpoints and convergence rather than a combinatoric one. 
 The use of a probability distribution 
 for measuring distance between voting rules 
 is crucial in that respect. 
% The results   
% put forward 
% the novel perspective on Arrow's theorem that 
% desirable election mechanisms necessarily converge towards a dictatorship, 
% where the convergence is with respect to 
% the process of transferring votes to the most ``powerful'' voter. 
% The results   
%suggest 
%that Arrow's theorem can be interpreted as saying  that 
%  election mechanisms necessarily converge towards a dictatorship, 
% where the convergence is with respect to 
% the process of transferring votes to the most ``powerful'' voter. 
%% The formulaton does not make clear which process of convergence you mean.
%%Please reformulate to make clear that the convergence is wrt the process of iterating \Phi,      ie wrt the process of 
%%transferring votes to the most influential voter. 
Our 
%modus operandi 
employment of fixpoints
% of working 
connects Arrow's theorem to the 
 literature of mathematical economics, where 
 many equilibrium concepts, such as for example 
 Nash equilibrium~\cite{nash1950equilibrium}, arise as a fixpoint.

We finish the introduction by discussing related work. 
 A vast number of proofs have appeared since Arrow's first demonstration 
 of the impossibility theorem.
Arrow's original proof  
%which interestingly contained a minor error~\cite{blau1957existence},
proceeded by showing the existence of a ``decisive'' voter. 
This approach was then refined by others, such as Blau~\cite{blau1972direct} 
and Kirman and Sondermann~\cite{kirman1972arrow}, who used ultrafilters.
%involves defining semi-decisive groups, proving a ?field expansion lemma? establishing that semi- decisive groups are decisive, proving a ?group contraction lemma? establishing that a decisive group contains a proper decisive group, and concluding that there is a single decisive individual.
Barbera~\cite{barbera1980pivotal}  replaced  the notion of a decisive voter with the weaker notion of a ``pivotal'' voter.
In later work, Geanakoplos~\cite{geanakoplos2005three} and Reny~\cite{reny2001arrow} sharpened the latter approach 
by introducing the concept of ``extremely pivotal''  voter 
in order to obtain shorter proofs.
%~\cite{geanakoplos2005three,reny2001arrow}.
%proceeded in two steps: showing the existence of a decisive voter, and then showing that a decisive voter is a dictator. 
%Barbera replaced the decisive voter with the weaker notion of a pivotal voter, thereby shortening the first step, but complicating the second step. 
%
%
%I give three brief proofs, all of which turn on replacing the decisive/pivotal voter with an extremely pivotal voter (a voter who by unilaterally changing his vote can move some alternative from the bottom of the social ranking to the top), thereby simplifying both steps in Arrow's proof.
 Even though the aforementioned proofs are all distinct, 
% and indeed involve different notions (such as decisiveness and pivotalness),  
they all are 
% at heart 
% in essence 
 of a similar, combinatorial nature: 
 One proves results about properties such as decisiveness or pivotalness by defining and  manipulating profiles in a
 clever
%  ``smart'' 
  way that  
%  precisely
 allows for the desirable properties
 to be taken advantage of. 
%This is too informal. Be  more concrete/specific.
 To our knowledge, Kalai's~\cite{kalai2002fourier} proof of Arrow's theorem  based on Fourier analysis on the Boolean cube was the first approach to 
 take a 
% radically
  different stance.
% steer clear 
% of such combinatorial methods. 
 Kalai's core idea was to calculate the probability of having a Condorcet cycle under 
some  probability distribution. 
One advantage of this approach is that, by using 
a theorem from Boolean analysis by Friedgut, Kalai, and Naor~\cite{friedgut2002boolean}, it produces  a robust, quantitative version of the theorem, in the following sense:
%This quantitative variant of Arrow's theorem 
%states, 
%Roughly speaking, Kalai showed that 
The more one seeks to avoid Condorcet's paradox, the more the election scheme will look like a dictator.
Our approach is similar to Kalai's in that 
we use quantitative methods, but it differs from it 
as we prove the classical 
version of Arrow's theorem 
and not a quantitative one.

%Unlike Kalai's approach, we
%Roughly 
%which shows that being close to 
%Subsequent research has yielded similar quantitative analogues 
%of the Gibbard-Satterthwaite theorem
%%%%%%%%%%%%%%%%%   RELATED WORK STUFF

 \paragraph{Contents of this paper.}
In Section~\ref{sec-prelim}, we give a brief overview of the basic formal concepts 
that are used in this paper.
We also formulate Arrow's theorem in a precise way. 
%In Section~\ref{sec-newproof}, we first 
%give an outline of our proof strategy and put forward the subresults 
%that need to be achieved to that end.
In Section~\ref{sec-newproof}, we present our fixpoint-based proof of Arrow's theorem.
%In the subsequent subsections we then prove 
%each of those subresults. 
%This will conclude our proof of Arrow's theorem.
%In Section 3, we present our fixpoint-based proof of Arrow's theorem.
Finally, we conclude and discuss future work in
Section~\ref{sec-conclusion}. 
The proofs that are not given in the main body of the text 
can be found in the appendix.

%--------------
%We recall Arrow's famous impossibility theorem~\cite{arrowjko} from social choice theory: if a voting rule for a finite number of voters and at least three candidates satisfies the Pareto property as well as independence of irrelevant alternatives, then that voting rule must be a dictator. 
%
%There have been numerous proofs of this result. 
%We attempt a novel approach: to prove Arrow's theorem using a fixpoint argument, namely using Banach's fixpoint theorem for complete metric spaces. The basic idea is to see dictators as ``stable elements'' of a process, and Arrow's result as saying that satisfaction of certain desirable properties necessarily leads  (converges) to a dictatorship. 

\section{Preliminaries}		\label{sec-prelim}

In this section we introduce all basic notions that will be used later.

\subsection{Arrovian Framework for Social Choice}	\label{subsec-prelimsocch}

In this subsection we recall the basic Arrovian framework, going back to Arrow's 
original work~\cite{arrowjko}.  
%Throughout, a set $A$  is assumed
%This

%\begin{definition}	\label{def-LA}
%%Given a set $A$, we denote with $\mathcal{L}(A)$ the set of strict total orders on $A$.
%Given a set $A$,  let $\mathcal{L}(A)$ be the set of strict total orders on $A$.
%%Let $A$ be a set.
%%Then $\mathcal{L} ( A )$ is the set of strict total orders on $A$.
%\end{definition}

We assume that the $n$ voters are linearly ordered, 
so without loss of generality 
we may suppose that the set of voters is
  $\{ 1 , 2 , \ldots , n \}$
  with the natural ordering. 
Given a set $A$, we  let $\mathcal{L}(A)$ be the set of strict total orders on $A$.
For our purposes, $A$ will be the set of candidates of the election, 
which we shall always  assume to be finite and at least three. 
We can identify an element of $\mathcal{L} ( A )$  with a complete listing of the 
elements of $A$ in order of preference.
If $\ell \in \LA$ and $a,b \in A$, we also write $a \, \ell \, b$ to mean that $(a,b) \in \ell$. 
%\submission{}{
%E.g., if $A = \{ a,b,c,d \}$, then
%$b \quad d \quad c \quad a$
%is an element of $\mathcal{L} ( A )$, with the interpretation that 
%candidate $b$ is most preferred and $a$  least.
%}
We shall use $x = (x_1 , x_2 , \ldots , x_n)$ to refer to an element of $\LA^n$. 
Such an element is usually called a \emph{profile} in social choice theory.
The element $x_i$ is the \emph{individual preference} of voter $i$. 

\begin{definition}	\label{def-votingrule}
A  \emph{voting rule} for $n$ {voters} and set of candidates $A$ 
is a map $\LA^n \to \LA $.
\end{definition}

Here the interpretation is that if the preferences of the $n$ voters are respectively $x_1 , x_2 , \ldots , x_n \in \LA$, then the outcome of the election under the voting rule $f$ is $f(x_1 , x_2 , \ldots , x_n ) \in \LA$. 
The latter is the so-called \textit{social choice} 
given how the electorate voted. 

%\submission{}{
A particularly bad type of voting rule is the following:
%}

\begin{definition}	\label{def-dicts}
The $n$ projection maps $\LA^n \to \LA$ 
are called \emph{dictatorships}. 
The $i$-th dictator is denoted as $\Dict_i$, where $i= 1 , 2, \ldots , n$.
We furthermore define $\DICT^n = \{ \Dict_i \mid i= 1 , 2, \ldots , n \}$.
\end{definition}

%\submission{}{
Note that $\Dict_i ( x )  = \,  x_i$ for all $ x  = ( x_1 , x_2 , \ldots , x_n) \in \LA^n$. 
%}
We write $\DICT^n$ shorthand as $\DICT$.
% if $n$ is assumed. 

\begin{definition}	\label{def-props}
Let $f \colon \LA^n \to \LA $ be a voting rule for $n$ voters and set of candidates $A$.
We say that $f$ satisfies the
\begin{itemize}
\item \emph{Pareto property} if for all $a,b \in A$ and 
all $x  = ( x_1 , x_2 , \ldots , x_n) \in \LA^n$,
if
$a \, x_i \, b$ for all 
$i$,
%$i=1 ,2 , \ldots, n$, 
then
$a \, {f( x )} \, b$.

\item \emph{independence of irrelevant alternatives (IIA) property} in case 
for all $a , b \in A$ and for  all profiles 
$x = ( x_1 , x_2 , \ldots , x_n) \in \LA^n$ 
and 
$x'  = ( x_1' , x_2' , \ldots , x_n' ) \in \LA^n$,
if ($a \, x_i \, b$ if{f} $a \, x_i' \, b$) for all 
$i$,
%$i=1 ,2 , \ldots,  n$,
then ($a \, f(x) \, b $ if{f} $a \, f( x' ) \, b $). 
\end{itemize}
\end{definition}

%\submission{
%%The Pareto property says that 
%% for any candidates $x$ and $y$, if the \emph{whole} electorate prefers $x$ to $y$, then also the social choice ought to prefer $x$ to $y$. 
%The Pareto property says that 
%if the electorate is unanimous about any two candidates, then 
%the social choice should respect that. 
%}
%{
The Pareto property speaks for itself: It states that for any candidates $a$ and $b$, if the \emph{whole} electorate prefers $a$ to $b$, then surely also in the social choice $a$ ought to be preferred to $b$. 
%}
%\submission{
%The IIA property says that 
%the relative social ranking
%of two alternatives solely depends on their relative individual rankings.
%}{
The independence of irrelevant alternatives property is a bit more dif{f}icult to grasp. 
It says that  the relative social ranking
of two alternatives only depends on their relative individual rankings.
In that sense there should be no dependence on any other alternative (hence ``irrelevant''). 
%}

\begin{definition}	\label{def-PIIA}
 We let $\PIIA^n =  \{ f \colon \LA^n \to \LA \mid f \mbox{ satisfies Pareto and IIA} \}$.
\end{definition}

We write $\PIIA^n$ shorthand as $\PIIA$. 
Note that $ \DICT^n \subseteq \PIIA^n$.

Arrow set out to investigate whether there was a 
non-dictatorial voting rule satisfying both the Pareto condition as well as IIA. 
The answer turned out to be negative. 
%See 
%Theorem~\ref{thm-arrow} below for the formal statement 
%of Arrow's contrivance. 

%\begin{theorem}[Arrow]	\label{thm-arrow}
%A voting rule for at least three candidates that satisfies 
%the Pareto property and IIA must be a dictatorship. 
%\end{theorem}

\begin{thom}[Arrow]	\label{thm-arrow}
A voting rule for at least three candidates that satisfies the Pareto property and IIA must be a dictatorship, i.e., $\PIIA^n  = \DICT^n$ for all $n \geq 1$. 
\end{thom}

%\subsection{Extending Metrics to the Quotient Space}	\label{subsec-pseudometric}
\subsection{Metric Space Basics}	\label{subsec-pseudometric}

 We assume basic knowledge of metric  spaces such as introduced in, e.g.,~\cite{arhangel?skij1990general}. 
 Here we just recall a few basic definitions.
 A \emph{metric space} $(X,d)$ is a set $X$ 
 equipped with a metric.
% , where 
%% where a
%%  A metric space (X, dX ) is a set X equipped
%%Recall that a 
%a  \emph{pseudometric} is a metric for which the distance between two distinct points can be zero. 

A crucial element in our proof is Banach's fixpoint theorem~\cite{banach1922operations}. 
We recall that
a function $F \colon X\to X$ on a metric space $X$ is \textit{contractive}
if  there is a $C<1$ such that 
we have $d ( F(x_1) , F(x_2) ) \leq C \cdot d (x_1 , x_2)$ 
for all $x_1, x_2 \in X$. 
A \textit{fixpoint} of $F$ is an element $x^*$ 
such that $F(x^*) = x^*$. 
For any map $F \colon X \to X$, by $F^{(n)}$ we mean the 
$n$-fold composition $F \o \cdots \o F$. 

\begin{theorem} [Banach Fixpoint Theorem]	\label{theorem-Banach}
Let $(X,d)$ be a non-empty complete metric space.
  If $F \colon X\to X$ is contractive
%   (i.e., there is $C<1$ such that 
%  $d ( F(x_1) , F(x_2) ) \leq C \cdot d (x_1 , x_2)$ for all $x_1, x_2 \in X$),
then
$F$ has a unique fixpoint $x^*$.
%That is,  $F(x^*)  = x^*$.
For all $x\in X$, $x^* = \lim_{n\to\infty} F^{(n)}(x)$.
\end{theorem} 

%where $F^{(n)}$ is $n$-fold composition. 

When $X$ is a compact metric space 
 it suffices to show that 
$d(F(x_1) , F(x_2 )) < d(x_1, x_2) $ for all $x_1 , x_2 \in X$ 
to conclude that $F$ is a contraction. 
This is in particular true  if  $X$ is finite. 
Recall that if there is an $n$ such that $F^{(n)}$ is a contraction, then 
$F$ has a unique fixpoint (see, e.g.,~\cite{pata2014fixed}). 
We shall use this fact later.

Let $(X,d)$ be any metric space 
for which on the underlying set $X$  an equivalence
relation $\sim$ is defined. 
For reasons that will become clear later, 
we wish to extend the metric on $X$ to a metric on $X/  \hspace{-0.1cm} \sim$. 
In general, the following construction exists (see, e.g.,~\cite{cagliari2015natural} for the details). 
%Let $(X,d)$ be any \emphi{pseudo}metric space, 
%meaning that there might be distinct points with distance zero. 
A \emph{chain} $C$ between points $x \in X$ and $y \in X$  
%, denoted as $C$, 
is a sequence of points 
$ x = a_0 \sim b_0, a_1 \sim b_1 , \ldots , a_n \sim b_n = y$, 
and the \emph{length} of such
a  chain 
%$C$ 
is defined as 
$\mbox{length}(C) = \sum_{i=0}^{n-1} d(b_i , a_{i+1}) $
% (this sum is understood to be  zero when $n = 0$). 
 (this sum is  understood to be zero when $n = 0$). 
Define
\begin{equation}	\label{ref-equation-for-dsim}
d_\sim ( [  x ] , [  y ] ) = 
\inf \{ \mbox{length}(C) \mid \mbox{$C$ is a chain between $x$ and $y$} \}.
\end{equation}
It is easy to see that $d_\sim$ is well-defined and satisfies all 
axioms of a pseudometric (meaning that the distance between two distinct points can be zero). 
Note though that $d_\sim ( [  x ] , [  y ] ) = 0  \Rightarrow [  x ] = [  y ]$ is not generally true.
%, not even if $d$ were a metric. 
However, in case $d$ 
%is a metric and furthermore 
is such that the infimum is always attained (which happens in particular in case $X$ is finite), 
then
% it is not hard to see that
$d_\sim$ is, in fact, a metric:

\begin{lemma}	\label{lem-techprmetr}
Let $(X,d)$ be a metric space and $\sim$ an equivalence relation on $X$. 
If the infimum in $d_\sim$ is always attained, then $d_\sim$ is a metric on $X/  \hspace{-0.1cm} \sim$.
\end{lemma}
\vproofspace
\begin{proof}
Indeed, if 
$d_\sim ( [x] , [y] ) = 0  = d_\sim ( [x] , [y] ) = \mbox{length} (C)$ for some chain $C$ between
$x$ and $y$ given by 
$ x = a_0 \sim b_0, a_1 \sim b_1 , \ldots , a_n \sim b_n = y$, 
then $ d(b_i , a_{i+1})=0$ for all $i$.
Since $d$ is a metric, we get
$b_i = a_{i+1}$ for all $i$. 
Thus,  $x=a_0 \sim a_1 \sim \cdots \sim a_n \sim y$, which 
implies $x \sim y$ by transitivity. Hence, $[x] = [y]$.
\end{proof}

%In our case, we assume the distribution $D$ (from Definition~\ref{def-spaceset}) to have full support,
%so that by Proposition~\ref{prop-probismetric} we know that 
%$d_D$ is a metric on $M$. 
%Since $M$ is finite, it follows from the above that
%$d_\sim$ is a metric on $\MExt= M / \hspace{-0.1cm} \sim$. 
Nonetheless, the defining formula for $d_\sim$ is difficult to work with. 
%, so we  would like to find out if there is an easier formula for it. 
The following result  gives a sufficient 
condition for the formula to reduce to an easier one. 
It uses the notion of action and orbit under a group action (see, e.g., \cite{rotman2012introduction}).
A \emph{group action} of a group $G$ on a set $X$ 
is a map $\phi \colon G \times X \to X$ such that 
$\phi(e, x)=x$ for all $x \in X$, where $e$ is the
identity element of $G$, and 
$\phi ( g , \phi ( g' , x ) ) = \phi ( g g' , x )$ for all 
$g,g' \in G$ and $x \in X$. 
The \emph{orbit} of an 
%element 
$x \in X$ under 
the group action $\phi$ 
is the set 
$ \{ \phi ( g , x ) \mid g \in G \}$. 
We also recall that the set of isometries on a metric space forms a group, 
with function composition as group law. 
With a group of isometries we mean a subgroup of that group. 

\begin{proposition}		\label{prop-genpsmetr}
 If $(X,d)$ is a
metric space endowed with an equivalence relation $\sim$ 
where the equivalence classes are the orbits 
of the action of a group of isometries on $(X,d)$, then 
it holds   for all $[  x ] , [  y ] \in X / \hspace{-0.1cm} \sim$ that 
$d_\sim ( [  x ]  , [  y ]  ) =  
\inf \{ d ( x' , y' ) \mid x' \in [  x ] , y' \in [  y ] \}$.
%\begin{itemize}
%\item[(1)] $ d_\sim ( [  x ]  , [  y ]  ) =  
%\inf \{ d ( x' , y' ) \mid x' \in [  x ] , y' \in [  y ] \}  $  for all $[  x ] , [  y ] \in X / \hspace{-0.1cm} \sim$.  
%\item[(2)]  The topology induced by the quotient pseudometric 
%equals
%%coincides with 
%the quotient topology.
%\end{itemize}
\end{proposition}
\vproofspace
\begin{proof}
This is  part of Theorem 2.1 from~\cite{cagliari2015natural}.
\end{proof}

%We recall that
%a function $f \colon X\to X$ on a metric space $X$ is \textit{contractive}
%if  there is a $C<1$ such that for all $x_1, x_2 \in X$,
%we have $d_X ( f(x_1) , f(x_2) ) \leq C \cdot d_X (x_1 , x_2)$. 
%A \textit{fixpoint} of $f$ is an element $x^*$ 
%such that $f(x^*) = x^*$. 

%Theorem 2.1 from~\cite{cagliari2015natural} states that if $(X,d)$ is a
%pseudometric space endowed with an equivalence relation $\sim$ 
%where the equivalence classes are the orbits 
%of the action of a group of isometries on $(X,d)$, then 
%\begin{itemize}
%\item[(1)] $ d_\sim ( [  x ]  , [  y ]  ) =  
%\inf \{ d ( x' , y' ) \mid x' \in [  x ] , y' \in [  y ] \}  $  for all $[  x ] , [  y ] \in X / \hspace{-0.1cm} \sim$.  
%\item[(2)]  The topology induced by the quotient pseudometric coincides with the quotient topology.
%\end{itemize}

\section{Fixpoint Argument Proof}		\label{sec-newproof}

In this section, we give the new proof of Arrow's theorem. 
The general idea is to define a contraction that has the set of dictators as its unique fixpoint.  

\subsection{Strategy and Outline}	\label{subsec-proofstrat}

We wish to use Banach's fixpoint theorem.
%~\cite{banach1922operations}. 
In order to do so, we need to view our voting rules 
as part of some metric space. 
For technical reasons, we shall focus on voting rules that satisfy the \emphi{Pareto} property. Thus,  
%Much of the theory
we let $M$ be the set of all voting rules $ \LA^n \to \LA$ 
that are Pareto. 

The idea to view voting rules as a metric space
where the metric is based 
on a distribution is inspired by earlier work 
from the Boolean analysis approach to 
social choice theory, such as Mossel~\cite{mossel2012quantitativeor}. 
Viewing the distance between voting rules as the 
probability that they produce a different outcome 
under a given distribution has the
advantage that profiles can be considered 
synchronously. 
By framing voting in terms of probabilities, this naturally puts us 
on the way to a quantitative stance.

%\begin{definition}	\label{def-spaceset}
%Let $M$ be the set of all voting rules $ \LA^n \to \LA$ 
%that are Pareto. 
%\end{definition}

\begin{definition}	\label{def-metriconm}
Given a probability distribution $\Dis$ on $\LA^n$, we define 
a map $d_{\Dis} \colon M \times M \to \reals$ by
$d_{\Dis} (f,g) = \Prob_{x \sim \Dis} [ f(x) \neq g(x) ]. $
\end{definition}

That is, $d_{\Dis} (f,g)$ is the probability under $\Dis$ 
that $f$ and $g$ produce a different outcome. 
For any set $X$, by $1_X$ we mean the indicator function on $X$.
Note that  
$$d_{\Dis} (f,g) 
= \Prob_{x \sim \Dis} [ f(x) \neq g(x) ]
= \sum_{x \in \LA^n} \Dis (x) \, 1_{f(x) \neq g(x)}.
$$

Throughout this paper, $\Dis$ is a distribution on $\LA^n$, 
but   for brevity we often simply say  that ``$\Dis$ is a distribution''.
% for short.
For example, taking $\Dis$ to be the uniform distribution 
is known in the literature as the \emph{impartial culture assumption}~\cite{garman1968paradox}.
%Pr[ ?]  =  \sum_x D(x) \cdot 1_{?}    and introduce indicator 
%\begin{proposition}	\label{prop-probismetric}
%If $D$ is a probability distribution on $\LA^n$, 
%it holds that $d_D$ is a pseudometric on $M$. 
%If moreover $D$ has full support, then $d_D$ is a  metric on $M$.
%\end{proposition}
%The idea to view voting rules as a metric space
%where the metric is based 
%on a distribution is inspired by earlier work 
%from the Boolean analysis approach to 
%social choice theory, such as Mossel~\cite{mossel2012quantitativeor}. 
%Viewing the distance between voting rules as the 
%probability that they produce a different outcome 
%under a given distribution has the
%advantage that profiles can be considered 
%synchronously. 
%By framing voting in terms of probabilities, this naturally puts us 
%on the way to a quantitative stance.

Under a mild assumption on $\Dis$
we obtain our desired metric space.

\begin{proposition}	\label{prop-probismetric}
If $\Dis$ is a probability distribution on $\LA^n$ 
%it holds that $d_D$ is a pseudometric on $M$. 
with full support, then $d_{\Dis}$ is a  metric on $M$.
\end{proposition}
\vproofspace
%\submission{}{
\begin{proof}
Clearly $d_{\Dis} (f,f) = 0$ for all $f \in M$. 
Since 
%by assumption 
$\Dis$ has full support, 
 $d_{\Dis} (f,g) = \Prob_{x \sim \Dis} [ f(x) \neq g(x) ] = 0$ implies  $f=g$.
Symmetry follows immediately. 
To prove the triangle inequality, let $f,g,h \in M$. 
Note that for any $x$, 
$f(x) \neq h(x) $ implies $f(x) \neq g(x)$ or $g(x) \neq h(x)$. 
Thus, 
$$ \Prob_{x \sim \Dis} [ f(x) \neq h(x) ] 
\leq  \Prob_{x \sim \Dis} [ f(x) \neq g(x)  \mbox{ or }  g(x) \neq h(x)  ], $$
and the latter is at most  $ \Prob_{x \sim \Dis} [ f(x) \neq g(x) ] +   \Prob_{x \sim \Dis} [  g(x) \neq h(x)  ] $.
\end{proof}
%}

%Arrow's theorem states that $\DICT = \PIIA$, i.e., 
%$$ \{ \Dict_i \mid i = 1 , 2 , \ldots , n \} = \{ f \mid f \mbox{ satisfies Pareto and IIA} \}, $$
%for any $n $. 
%So that is what we need to show.

%The general proof idea is to define a contraction
%$\Phi \colon M \to M$ 
%that has a dictator(s) as (unique) fixed point. 
%The contraction we have in mind is as follows. 

Note that the metric space $(M, d_{\Dis})$ is complete and compact for any $\Dis$, since it is finite.
%The idea to view voting rules as a metric space
%where the metric is based 
%on a distribution is inspired by earlier work 
%from the Boolean analysis approach to 
%social choice theory, such as Mossel~\cite{mossel2012quantitativeor}. 
%Viewing the distance between voting rules as the 
%probability that they produce a different outcome 
%under a given distribution has the
%advantage that profiles can be considered 
%synchronously. 
%By framing voting in terms of probabilities, this naturally puts us 
%on the way to a quantitative stance.

%It is natural to see voting rules in terms of distance

%(M,_d_D)  is complete (because finite) for any D.

The proof idea is to define a contraction  that has the set of dictators as its unique fixpoint. We start by defining a map $\Phi$ on $M$. 
Essentially, given a voting rule $f$,  we will define $\Phi (f)$ as the voting rule in which the  ``least forceful'' voter's vote is replaced by the ``most forceful'' voter's vote. 
This idea is inspired by, though different from,
the notion of \emph{influence} 
that is well-studied in Boolean analysis~\cite{o2014analysis}. 
The \emphi{influence} of the $i$-th voter is defined as  the probability
that the $i$-th vote affects the outcome (assuming 
there are only two candidates).
In political science and voting theory, 
influence also goes by the name \emph{voting power}. 
%which was independently invented by several people, 
Voting power can be thought of as the ability of a legislator, by his vote, to affect the passage or defeat of a measure~\cite{banzhaf1964weighted}. 

To make the aforementioned idea formal, we define what we mean by the force of a voter.

 \begin{definition}	\label{def-infl}
Given a probability distribution $\Dis$ on $\LA^n$ and voter $i$, we define 
the \emph{force} of voter $i$ on voting rule $f$ under distribution $\Dis$ as 
$$ \Inf_{\hspace{-0.1cm} \Dis}^i [ f ] = \Prob_{x \sim \Dis} [ f(x) = x_i ]. $$  
The set of \emph{most forceful} voters is
$ \MostInf_{\hspace{-0.00cm}\Dis} [ f ] = \argmax_{i \in \{1,2, \ldots , n\}} \Inf_{\hspace{-0.1cm}\Dis}^i [ f ].$ \\
The set of  \emph{least forceful} voters is
$ \LeastInf_{\hspace{-0.00cm}\Dis} [ f ] = \argmin_{i \in \{ 1,2, \ldots , n\}} \Inf_{\hspace{-0.1cm}\Dis}^i [ f ].$
\end{definition}

%Throughout this paper, $D$ is always a distribution on $\LA^n$, 
%but we often simply say for brevity that ``$D$ is a distribution''.
%% for short.
%For example, taking $D$ to be the uniform distribution 
%is known in the literature as the \emph{impartial culture assumption}~\cite{garman1968paradox}.
%, as follows. 

%We use the concept of force to 
%make the aforementioned idea precise. 
The notion of force 
was first introduced in~\cite{Feys:Thesis:2015}.

%Point out the choice for the 
%min/first voter is arbitrary, and that it is a technicality
%to endure the votes are transferred to one specific voter. 

 \begin{definition}	\label{def-contrctphi}
Let $\Dis$ be a given distribution with full support.
%on $\LA^n$. 
Given a voting rule $f \in M$, we define a new voting rule $\PhiD (f)\in M$ by 
$ \PhiD (f) ( x_1 , \ldots , x_n ) = f( y_1 , \ldots , y_n )$,
%$ \Phi_{\sim}^D (f) ( x_1 , \ldots , x_n ) = f( y_1 , \ldots , y_n )$,
%$ \Phi^{\sim} (f) ( x_1 , \ldots , x_n ) = f( y_1 , \ldots , y_n )$,
%$ \PhiD (f) ( x_1 , \ldots , x_n ) = f( y_1 , \ldots , y_n )$,
where $$
y_i = \begin{cases} x_i &\mbox{if } i \not\in \LeastInf_{\Dis} [f],  \\ 
x_{ \min \MostInf_{\Dis} [f]} & \mbox{else}.  \end{cases}
$$
This defines a map  $\PhiD \colon M \to M$, also written 
for short as $\Phi$ if $\Dis$ is understood from the context.
 \end{definition}

 So what happens when passing from $f$ to $\Phi (f)$ is that all 
  voters with least force 
have their vote replaced by the vote of the
 {first} voter that has maximal force. 
 Note that $\Phi ( f )$ is Pareto if $f$ is Pareto, 
 so $\Phi$ is well-defined on $M$.   
 The choice to consistently 
pick specifically the first voter in Definition~\ref{def-contrctphi} is arbitrary. 
We might as well always pick the last voter, for instance.
Our choice is merely a technicality to 
ensure that the votes are transferred to one specific voter.

% \begin{definition}	\label{def-contrctphi}
%Let $D$ be a given distribution.
%Given a voting rule $f \in M$, we define a new voting rule $\Phi (f)\in M$ by 
%$ \PhiD (f) ( x_1 , \ldots , x_n ) = f( y_1 , \ldots , y_n )$,
%where $$
%y_i = \begin{cases} x_i &\mbox{if } i \not\in \LeastInf_D [f],  \\ 
%x_{  \MostInf_D [f]} & \mbox{else}.  \end{cases}
%$$
%Here, $x_{  \MostInf_D [f]} $ is any element of $\{ x_i \mid i \in  \MostInf_D [f] \}$. 
%For any such  choice, this defines a map  $\PhiD \colon M \to M$, also written 
%for short as $\Phi$ if $D$ is understood.
% \end{definition}
 
%If the distribution $D$ is assumed, we also simply write $\Phi$ instead of $\PhiD$.

\begin{proposition}	\label{prop-dictisfixpointerst}
For all $i$ we have  $\Phi ( \Dict_i ) = \Dict_i$.
\end{proposition}
\vproofspace
\begin{proof}
%We show that $\Phi ( \Dict_i  ) \sim \Dict_i $ for any $i$.
Voter $i$ is the unique most forceful voter of $\Dict_i$.
It  is irrelevant how others vote, and we have 
 $\Phi ( \Dict_i  )  (x_1 , \ldots , x_n )   = x_i$ for all $(x_1, \ldots , x_n)$. Hence, 
 $\Phi ( \Dict_i  ) = \Dict_i$. 
% This implies the equality 
% $ \Phi_\sim( \DICT ) = \Phi_\sim ( [ \Dict_i ] )
% = [ \Phi ( \Dict_i ) ]
% = [ \Dict_i ] = \DICT$, 
% where we used Proposition~\ref{prop-dictformseqclass} several times.
\end{proof}

Note that $\Phi$ cannot be a contraction, since \emph{every} dictator is a fixed point of $\Phi$, contradicting the uniqueness of the fixed point in Banach's fixpoint theorem.
The idea is therefore to consider all dictators to be ``the same'', i.e., 
to consider them to be equivalent under some equivalence relation. 
%Thus, we have to find an equivalence relation $\sim$ on $M$ such that the equivalence class of a dictator is the set of dictators. 

Given an equivalence relation $\sim$, we write $[  x ]$ for the equivalence class of $x$ under $\sim$. 
We shall write $\MExt$ for $M / \hspace{-0.1cm} \sim$ throughout. 
%Our strategy is the following.

\vspace{0.2cm}

The remainder of the proof is organized as follows.
We define an equivalence relation  $\sim$ using a notion of permutation-invariance of voting rules as well as our notion of force. We then show that $\DICT$ is an 
equivalence class of $\sim$. 
The use of permutations gives rise to a notion of permutation-invariance for distributions.
We then show that for all distributions $\Dis$ on $\LA^n$ that have full support and are permutation-invariant, 
the following hold. 
\begin{enumerate}
\item[(i)]  The map 
$\Phi^{\Dis} \colon M^{\Dis} \to M^{\Dis}$  is compatible with $\sim$, where $M^{\Dis}$ is the metric space resulting from $\Dis$,
and hence extends to a map $\Phi^{\Dis}_\sim \colon M^{\Dis}_\sim \to M^{\Dis}_\sim$.
\item[(ii)] The map $\Phi^{\Dis}_\sim$ has $\DICT$ as unique fixpoint.
\end{enumerate}
Finally, we prove Arrow's theorem by showing the existence of a 
concrete distribution $\Dis$ satisfying the abovementioned requirements.

%
%%I think it is better to omit the last (rather technical)  bullet here, and incorporate it into the actual proof. See my comments later.
%
%
%
%If all of this can be achieved, then Arrow's theorem follows. 
%Indeed, 
%if by contradiction we had $\PIIA^{n-1} \neq \DICT^{n-1}$
%for some $n \geq 4$,   
%from (c) we would get an $ f \in  \PIIA \setminus \DICT$ such that $\Phi ' ( [ f ] ) = [f]$. 
%By uniqueness of the fixpoint in (b), from (c) it follows that $[f] =\DICT$, 
%so $f \in [f ] = \DICT$. 
%Contradiction.
%Hence, 
%$\PIIA^{n-1} = \DICT^{n-1}$ for all $n \geq 4$. 

%i.e., Arrow's theorem. 

%So the question is: how can our desiderata be satisfied? 

\subsection{Equivalence Relation on $M$ and Metric on $\MExt$}	\label{subsec-satparta}

%%%%%%%
%DOESN"T WORK
%
%One idea would be to use the following well-known theorem:
%
%\begin{theorem}	\label{thm-helpthm}
%Let $d$ be a pseudometric on a set $M$. 
%If we define a relation  $\sim$ on $M$ by letting 
%$x \sim y$ if{f} $d(x,y)=0$, then $\sim$ is an equivalence relation. 
%Furthermore, $d'$ defined by $d( \overline x , \overline y ) = d(x,y)$ is well-defined and in fact a metric on $\MExt = M / \hspace{-0.1cm} \sim$. 
%\end{theorem}
%
%The idea would be to pick the probability distribution $D$ from Definition~\ref{def-spaceset} in such a way that the $\sim$ obtained by $x \sim y$ if{f} $d(x,y)=0$ (as in Theorem~\ref{thm-helpthm}) makes desideratum (1) true. 
%%%%%%%

We first have to find an appropriate equivalence relation on $M$. 
To do that, we use the following notion. 

 \begin{definition}	\label{def-pixf}
Given  a permutation $\pi  \colon n \to n$ of the voters, we 
define $\piv (x)$ as $( x_{\pi (1) } , \ldots , x_{\pi(n)} )$ for all 
tuples $x= (x_1 , \ldots  , x_n) \in \LA^n$.
If  $f$ is a voting rule, 
we define a new voting rule, written as $f \o \piv$, by
 $( f \o \piv ) (x) = f( \piv (x))$ for all $x \in \LA^n$.
%we define $f( \pi (x) )$ as $f( x_{\pi (1) } , \ldots , x_{\pi(n)} )$ for all $x \in \LA^n$. 
%This voting rule is written as $f \o \pi$.
\end{definition}

%In what follows,  $\pi$ is always
% assumed to be a bijection $\pi \colon n \to n$. 

A permutation $\pi \colon n \to n$ is in what follows 
often written without types simply as $\pi$.
Observe that  $\vec{\pi} \colon \LA^n \to \LA^n$ is a bijection because $\pi$ is a bijection.

 \begin{definition}	\label{def-preeqrel}
We define the relation $\sim$ on $M$ by letting $f  \sim  g$ if{f} 
(i) $f=g$, or (ii) there exists a permutation $\pi \colon n \to n$ such that $f = g \o \piv$ and $f$ has a unique voter with maximal force.
%at least one of the following conditions holds: 
%(1) $f=g$, or (2) 
%\begin{itemize}
%\item[(1)] $f=g$.
%\item[(2)] There exists a permutation $\pi \colon n \to n$ such that $f = g \o \piv$, and $f$ has a unique voter with maximal force.
%\submission{}{(i.e.,  
%$ f(x) = g ( \pi ( x ) ) $ for all $x \in \LA^n$)}
%\item[(2)] $\pi ( \FMI  ( \Phi ^ {(k)} (g) ) ) = \FMI ( \Phi ^ {(k)} (f) )$ for all $k\geq0 $. 
%\end{itemize}
\end{definition}

% \begin{definition}	\label{def-eqrel}
%For $f,g \in M$, let $f \sim g$ if{f} 
%there exists a permutation $\pi \colon n \to n$ such that $f = g \o \pi$\submission{.}{, i.e.,  
%$ f(x) = g ( \pi ( x ) ) $ for all $x \in \LA^n$.}
%\end{definition}

%\begin{proposition}	\label{prop-isreflsymm}
%$R$ is an equivalence relation on $M$.
%\end{proposition}
%
%\submission{}{
%\begin{proof}
%Clearly $R$ is reflexive since we can simply take the identity permutation. 
%If $f \, R \, g$, then $f = g \o \pi$ for some $\pi$.
%It is easy to see that then $g = f \o \pi^{-1}$, showing symmetry. 
%%Since $\sim$ is the transitive closure of $R$, the result follows. 
%To show transitivity, suppose $f \, R \, g $ and $g \, R \, h$. 
%Then, $f = g \o \pi$ and $ g = h \o \sigma$ for some  $\pi, \sigma$. 
%Thus, $f = (h \o \sigma ) \o \pi$, and it is clear that
%this implies that $f = h \o ( \sigma \o \pi )$, so $f \, R \, h$.  
%  
%\end{proof}
%}

% \begin{definition}	\label{def-eqrel}
%Let $\sim$ be the transitive closure of $R$.
%\end{definition}

In this way we obtain our desired equivalence relation. 

\begin{proposition}	\label{prop-iseqrel}  	\label{prop-isreflsymm}
The relation $\sim$ is an equivalence relation on $M$.
\end{proposition}
\vproofspace
%\submission{}{
\begin{proof}
Clearly $\sim$ is reflexive, as $= \; \subseteq \; \sim$. 
To show symmetry, let $f \sim g$, so  that
$f=g$, or $f = g \o \piv$ for some $\pi$ where $f$ has a unique maximal force voter. 
The former case $f=g$ is trivial. 
In the latter case, clearly $g = f \o \piv^{-1}$, and 
because $f$ has a unique maximal force voter, the relation 
$f = g \o \piv$ implies that the same holds for $g$.
Now to show transitivity, let $f \sim g$ and $g \sim h$. 
Thus, $f=g$, or $f = g \o \piv$ for some $\pi$ where $f$ has a unique maximal force voter. 
Also, $g = h$, or $g = h \o \tauv$ for some $\tau$ where $g$ has a unique maximal force voter.
There are four cases. 
We will only show two of them, as they are all easy.
 
Suppose $f= g$ and $g = h \o \tauv$ where $g$ has a unique maximal force voter. 
Then since $g$ has a unique maximal force voter, the same holds for $h$, 
so the relation $f = h \o \tauv$ implies that this is also true for $f$. 
If $f = g \o \piv$ and $g = h \o \tauv$,  
where both $f$ and $g$ have a unique maximal force voter, 
we have that  $f = h \o ( \tauv \o \piv )$.
%\begin{enumerate}
%\item $f=g$ and $g=h$. Then, $f=h$.
%\item $f= g$ and $g = h \o \tauv$ where $g$ has a unique maximal force voter. 
%Then since $g$ has a unique maximal force voter, the same holds for $h$, and 
%the relation $f = h \o \tauv$ hence implies that this is also true for $f$.
%\item $f = g \o \piv$ and $g=h$ where $f$ has a unique maximal force voter. Then $f = h \o \piv$, so the conclusion follows.
%\item $f = g \o \piv$ and $g = h \o \tauv$,  
%where both $f$ and $g$ have a unique maximal force voter. 
%Then, $f = h \o ( \tauv \o \piv )$.
%\end{enumerate}
%This completes the proof.
\end{proof}
%}

%%I think it would be good to point 
%%out that equivalence classes are of 
%%the following two types:
%%- If f does not have a unique MIV,
%%then [f] = {f}
%%- If f does have a unique MIV, then
%%  [f] = { f o \pi | \pi}?

Note that $\DICT \subseteq M$, so it makes sense to speak about $[ \Dict_i ]$, 
and that the equivalence classes of $\sim$ are of the
following two types:
\begin{itemize}
\item If $f$ does not have a unique maximal force voter, then $[f] = \{ f \}$.
\item If $f$ does have a unique maximal force voter, then 
$[f] = \{ f \o \piv \mid \pi \colon n \to n \}$. 
\end{itemize}

\begin{proposition}	\label{prop-dictformseqclass}
The set $\DICT$ is an equivalence class of  $\sim$.
%\submission{}{ the equivalence relation} $\sim$.
\end{proposition}
\vproofspace
\begin{proof}
  For any $i$, we show that 
we have $[ \Dict_i ] = \DICT$.
Let $f \in M$ be such that $f \sim \Dict_i$. 
If $f = \Dict_i$, surely $f \in \DICT$. 
Otherwise, we have that $f = \Dict_i \o \, \piv$ for some permutation $\pi$. 
Suppose $\pi (i ) = j$. 
A trivial computation then shows that $f = \Dict_j$, so $f \in \DICT$.
%It thus follows by the definition of $\sim$ and an easy induction 
%that $f \sim \Dict_i$ implies that $f$ is a dictator. 
%By the definition of $\sim$ and an easy induction 
%it follows that $f \sim \Dict_i$ implies that $f$ is a dictator. 
We now show that for any $j$,  
$\Dict_i \sim \Dict_j$.
Indeed, if $\pi$ is the permutation 
that switches $i$ and $j$ and is the identity elsewhere, 
then  $\Dict_i = \Dict_j \o \,  \piv$. 
Furthermore, it is clear that any dictator has 
a unique voter with maximal force.
\end{proof}

Let $\Dis$ be a distribution with full support such that $d = d_{\Dis}$ is a metric on $M$ (by Proposition~\ref{prop-probismetric}), and consider the quotient metric $d_\sim$ defined in~\eqref{ref-equation-for-dsim}
for the relation $\sim$
from Definition~\ref{def-preeqrel}.
Since $M$ is finite, we obtain from  Lemma~\ref{lem-techprmetr}
that $d_\sim$ is a metric on $M_\sim= M / \hspace{-0.1cm} \sim$.
However, as we saw in that same subsection, 
the defining formula for $d_\sim$ is convoluted, so 
we set out 
%to investigate whether we could 
to identify a subgroup of isometries that will allows us to
apply
Proposition~\ref{prop-genpsmetr}, as 
that would give us a more convenient formula to work with. 
To do that, we shall need the following notion. 

 \begin{definition}	\label{def-permuinvs}
A distribution $\Dis$ on $\LA^n$
 is \emph{$n$-permutation-invariant} if for all 
 $\pi \colon  n\to n$, $\Dis \circ \piv = \Dis$.
 \end{definition}

The condition $\Dis = \Dis \o \piv$ for all permutations $\pi$ 
demands that 
$ \Dis (x) = \Dis (x')$ whenever $x'$ and $x$ 
are rearrangements of one another. 
Mathematically, it ensures that 
the probability distribution $\Dis$ is well-defined up to permutation equivalence 
of profiles. 

%Given  a distribution $D$, we write $d_D$ for short as $d$ 
% if $D$ is understood from the context. 
%We wish to turn the metric space $(M, d)$ into a metric space $(M/  \hspace{-0.1cm} \sim , d_\sim)$. 
%In our case, we always assume the distribution $D$ 
%appearing in $d_D$ 
%%(from Definition~\ref{def-metriconm})
% to have full support,
%so that by Proposition~\ref{prop-probismetric}  
%we know that 
%$d_D$ is a metric on $M$. 
%Since $M$ is finite, it follows from
% the discussion in Subsection~\ref{subsec-pseudometric} that 
%$d_\sim$ is a metric on $\MExt= M / \hspace{-0.1cm} \sim$. 
%However, as we saw in that same subsection, 
%the defining formula for $d_\sim$ is convoluted, so 
%we set out to investigate whether we could apply
%Proposition~\ref{prop-genpsmetr} as 
%that would give us a more convenient formula to work with. 

For each permutation $\pi \colon n \to n$, let
$ J_\pi \colon M \to M \colon f \mapsto  f \o \piv. $
This map is  well-defined, as $f \o \piv$ is Pareto 
if $f$ is. 

\begin{proposition}	\label{prop-Jpiisisom}
Let  $\Dis$  be a distribution on $\LA^n$ 
with full support and such that $\Dis$  is $n$-permutation-invariant.
%(i.e., with full support and  with $D = D \o \piv$ for all $\pi$.
%\submission{with $D = D \o \piv$ for all permutations $\pi$.}{satisfying that
%$ D( x ) = D ( \piv ( x )   ) $
%for all permutations $\pi$ and all $x \in \LA^n$.}
Then 
$J_\pi$ is an isometry of $M$, for each $\pi$.
\end{proposition}
\vproofspace
\begin{proof}
From the bijectivity of $\pi$, it follows that $J_\pi$ is bijective. 
Let $\pi$ be a permutation 
and $f,g \in M$.
Then we need to show that $d_{\Dis} ( f, g ) = d_{\Dis} ( f \o \piv , g \o \piv )$. 

We calculate: 
\begin{eqnarray*}
\Prob_{x \sim \Dis} [ (f \o \piv ) (x) \neq ( g \o \piv ) (x) ] & = &
\sum_{x \in \LA^n}  \Dis ( x ) \cdot 1_{(f \o \piv) ( x ) \neq (g \o \piv) ( x ) }  \\ 
& = & \sum_{x \in \LA^n} \hspace{-0.0cm} \Dis ( \piv^{-1} (x) ) \vspace{-0.2cm} \cdot  \vspace{-0.2cm} 1_{(f \o \piv) ( \piv^{-1} (x) ) \neq (g \o \piv) ( \piv^{-1} (x) ) } \\
& = & \sum_{x \in \LA^n}  \Dis ( \piv^{-1} (x) ) \cdot 1_{f  ( x ) \neq g  ( x ) }.  
\end{eqnarray*}
Hence, 
$ \Prob_{x \sim \Dis} [ (f \o \piv ) (x) \neq ( g \o \piv ) (x) ] =
 \Prob_{x \sim \Dis \, \o \, \piv^{-1}} [ f (x) \neq  g  (x) ] $.
Note that  $\Dis \o \piv^{-1} \colon \LA^n \to [0,1]$ is indeed
a probability distribution on $\LA^n$. 
The proof is complete after observing that $\Dis \, \o \, \piv^{-1} = \Dis$.
\end{proof}

Let  $G = \{ J_\pi \mid \pi \mbox{ is a permutation} \}$. 
%We recall that the set of isometries on a metric space forms a group, 
%with function composition as group law. 
From Proposition~\ref{prop-Jpiisisom} 
it follows that $G$ is a subgroup of the group of isometries of $M$. 
We define a map 
%$ \phi \colon G \times M \to M$ by $\phi (g,f) = g(f). $
$ \cdot \colon G \times M \to M$ by 
$$ J_\pi \cdot f = 
\begin{cases} 
J_\pi (f)  & \mbox{if $f$  has a unique maximal force voter}, \\
f & \mbox{otherwise}.
\end{cases}
$$

\begin{proposition}	\label{prop-itsgroupaction}
The operation $\cdot$ is a group action of the group $G$ on $M$
and its orbits coincide with the equivalence classes under $\sim$.
\end{proposition}
\vproofspace
\begin{proof}
Note that $\mbox{id} \colon M \to M$ is the identity of $G$, and 
$\mbox{id} \cdot f  = f$ for all $f \in M$.
Also, $(g \o h ) ( f  ) = g ( h (f))$ for all $g,h \in G$ and $f \in  M$.
Indeed, 
this easily follows by making the case distinction 
whether  $f$ has a unique maximal force voter.

The orbit of an $f \in M$ under this group action is equal to
$ G \cdot f 
= \{ J_\pi \cdot f \mid \pi \}$.
Now, $J_\pi \cdot f$ is $ J_\pi (f) = f \o \piv$ if   $f$ has a unique maximal force voter, and 
$f $ otherwise. 
Hence, $ G \cdot f 
= \{ J_\pi \cdot f \mid \pi \} 
=  [ f ]$.
Thus, the orbits of the group action  coincide with the equivalence classes under $\sim$.
\end{proof}
  
%\begin{proof}
%Note that $\mbox{id} \colon M \to M$ is the identity of $G$, and 
%$\mbox{id} \cdot f  = f$ for all $f \in M$.
%Also, $(g \o h ) ( f  ) = g ( h (f))$ for all $g,h \in G$ and $f \in  M$.
%%} 
%Indeed, 
%this easily follows by making the case distinction 
%whether  $f$ has a unique maximal force voter.
%
%The orbit of an $f \in M$ under this group action is equal to
%$ G \cdot f 
%%=\{ \phi ( g , f ) \mid g \in G \} 
%= \{ J_\pi \cdot f \mid \pi \}$.
%Now, $J_\pi \cdot f$ is $ J_\pi (f) = f \o \piv$ if   $f$ has a unique maximal force voter, and 
%$f $ otherwise. 
%Hence, $ G \cdot f 
%%=\{ \phi ( g , f ) \mid g \in G \} 
%= \{ J_\pi \cdot f \mid \pi \} 
%=  [ f ]$.
%Thus, the orbits of the group action  coincide with the equivalence classes under $\sim$.
%\end{proof}

%It is clear that $\cdot$ is a group action of the group $G$ on $M$. 
%\submission{}{
%Indeed, note that $\mbox{id} \colon M \to M$ is the identity of $G$, and 
%$\mbox{id} ( f ) = f$ for all $f \in M$;
%also, $(g \o h ) ( f  ) = g ( h (f))$ for all $g,h \in G$ and $f \in  M$.} 
%Note that the orbit of an element $f \in M$ under this group action is equal to
%$ G \cdot f 
%%=\{ \phi ( g , f ) \mid g \in G \} 
%= \{ J_\pi (f) \mid \pi \} =
%\{ f \o \pi \mid  \pi  \}  =   [  f ]_R $.
%Thus, the orbits exactly coincide with the equivalence classes under $R$.}
%\submission{Theorem 2.1 from~\cite{cagliari2015natural} then implies the following.
%}{}

\begin{proposition}	\label{prop-formmetricreduces}
Let  $\Dis$  be a distribution with full support and 
{with $\Dis = \Dis \o \piv$ for all permutations $\pi$.}
%{satisfying that $ D( x ) = D ( \piv ( x )   ) $ for all permutations $\pi$ and all $x \in \LA^n$.}
Let $d_\sim$ be the metric on $M / \hspace{-0.1cm} \sim$ 
based on the metric $d = d_{\Dis}$ on $M$ (see Lemma~\ref{lem-techprmetr}).
Then for all $[  f ] , [  g ] \in M /  \hspace{-0.1cm} \sim$ we have
%\submission{}{we have }
%$$ d_\sim ( \overline f , \overline g ) =  
%\min \{ d ( f' , g' ) \mid f' \in \overline   f , g' \in \overline g \}.   $$
$ d_\sim ( [  f ] , [ g ] ) =  
\min \{ d ( f' , g' ) \mid f' \in [  f ]_\sim  , g' \in [  g ]_\sim  \}.   $
%\submission{Also, the topology  
%induced by the  metric $d_\sim$ coincides with the quotient topology.}{Furthermore, the topology  on $M /   \hspace{-0.1cm} \sim$
%induced by the quotient metric $d_\sim$ coincides with the quotient topology.}
\end{proposition}
%\submission{}{
\vproofspace
\begin{proof}
This follows from 
Proposition~\ref{prop-genpsmetr}, Proposition~\ref{prop-Jpiisisom}, and 
Proposition~\ref{prop-itsgroupaction}. 
%Theorem 2.1 from~\cite{cagliari2015natural} states that if $(X,d)$ is a
%pseudometric space endowed with an equivalence relation $\sim$ 
%where the equivalence classes are the orbits 
%of the action of a group of isometries on $(X,d)$, then 
%\begin{itemize}
%\item[(1)] $ d_\sim ( [  x ]  , [  y ]  ) =  
%\inf \{ d ( x' , y' ) \mid x' \in [  x ] , y' \in [  y ] \}  $  for all $[  x ] , [  y ] \in X / \hspace{-0.1cm} \sim$.  
%\item[(2)]  The topology induced by the quotient pseudometric coincides with the quotient topology.
%\end{itemize}
%The previous discussion shows that 
%the equivalence classes of $\sim$ are the orbits of the 
%group action $\cdot $ of the group of isometries $G$ on $M$, so the result follows. 
\end{proof}
%}

%\begin{proposition}	\label{prop-formmetricreduces}
%Let  $D$  be a distribution with full support and 
%\submission{with $D = D \o \pi$ for all permutations $\pi$.}{
%satisfying that
%$ D( x ) = D ( \pi ( x )   ) $
%for all permutations $\pi$ and all $x \in \LA^n$.}
%Let $d_\sim \colon M \times M \to \R$ be the following map on $M /  \hspace{-0.1cm} \sim$:
% for all $[  f ] , [  g ] \in M /  R$, 
%$$ d_\sim ( [  f ] , [ g ] ) =  
%\min \{ d ( f' , g' ) \mid f' \in [  f ]_\sim  , g' \in [  g ]_\sim  \}.   $$
%Then $d_\sim$ is a metric on $M$.
%\end{proposition}
%
%\submission{}{
%\begin{proof}
%The only nontrivial item to prove is the triangle inequality. 
%  
%\end{proof}
%}

%\subsection{Satisfying (b)}	\label{subsec-satpartb}

\subsection{Extending $\Phi$ to a Map with Unique Fixpoint} 	\label{subsec-satpartb}  \label{subsec-satpartc}

%\begin{proposition}	\label{prop-phiprimiswelldef}
%The map $\PhiExt$ from Definition~\ref{def-contrctphiprim} is well-defined. 
%\end{proposition}
%
%\submission{}{
%\begin{proof}
%Let $f,g \in M$ be such that $f \sim g$. 
%We need to show that $\Phi ( f ) \sim \Phi ( g )$. 
%Let $\pi$ be such that $f = g \o \pi$. 
%We will show that $\Phi ( f ) = \Phi ( g ) \o \pi$. 
%
%By what we just showed, we know that $ \Inf_D^i [ g ] =  \Inf_D^{\pi (i)} [ f ]$ for each $i$.     
%
%\end{proof}
%}

From now on, unless specifically mentioned otherwise, we shall always assume that 
 $\Dis$  is a distribution on $\LA^n$ with full support and 
with $\Dis = \Dis \o \piv$ for all permutations $\pi$, 
such that Proposition~\ref{prop-formmetricreduces} applies.
Such a $\Dis$ obviously exists, for instance the uniform distribution 
is an example, 
but one can easily obtain many  other examples 
simply by giving one representative of each permutation equivalence class 
of profiles (meaning profiles $x$ and $y$ are 
equivalent if{f} there is a permutation $\pi$ such that $\pi (x) = y$) 
a non-zero weight. 
Later we will exploit this property. 
%For such $D$,
% by 
%Proposition~\ref{prop-formmetricreduces} we have
%%we know that 
%%$d_\sim$ satisfies 
%$ d_\sim ( [  f ] , [ g ] ) =  
%\min \{ d ( f' , g' ) \mid f' \in [  f ]_\sim  , g' \in [  g ]_\sim  \}$. 
 
Our aim in this subsection is to extend the map $\Phi \colon M \to M$ to 
a map $\MExt \to \MExt$ where, as we recall, 
$\MExt = M /  \hspace{-0.1cm} \sim$. 
%We remind 
%%ourselves 
%that, although it is not written explicitly, 
%the metric $d_D$ on $M$ depends on the choice 
%of the distribution $D$.  
The following  is a technical result that we shall use later.

\begin{lemma}	\label{lem-techpirel}
If $f = g \o \piv$ for $\pi$, then $ \Inf_{\hspace{-0.1cm}\Dis}^i [ g ] =  \Inf_{\hspace{-0.1cm}\Dis}^{\pi (i)} [ f ]$ for each $i=1,2, \ldots ,n$.
\end{lemma}
\vproofspace
\begin{proof}
For any $i$, let $p_i \colon \LA^n \to \LA$ be the $i$-th projection map. 
Then we have
%\begin{eqnarray*}
% \Inf_{\hspace{-0.1cm}\Dis}^i [ g ] & = &  \Prob_{x \sim \Dis} [ g(x) = x_i ] \\
%& = & \sum_{x }  \Dis ( x ) \cdot 1_{g(x) = p_i (x ) }  \\ 
%& = &  \sum_{x }  \Dis ( \piv (x) ) \cdot 1_{g(\piv(x)) = p_i ( \piv (x))}.  
%\end{eqnarray*}
$$ \Inf_{\hspace{-0.1cm}\Dis}^i [ g ]  
 =   \Prob_{x \sim \Dis} [ g(x) = x_i ] 
=  \sum_{x }  \Dis ( x ) \cdot 1_{g(x) = p_i (x ) } 
 =   \sum_{x }  \Dis ( \piv (x) ) \cdot 1_{g(\piv(x)) = p_i ( \piv (x))}.  $$ 
Note that $p_i ( \piv (x) ) = x_{\pi (i )}$ 
for each $x$,
%$x \in \LA^n$, 
%and hence
so
$  \Inf_{\hspace{-0.1cm}\Dis}^i [ g ]  =  \sum_{x }  \Dis ( \piv (x) ) \cdot 1_{g(\piv(x)) = x_{\pi(i)} }$. 
By assumption, $\Dis \o \piv = \Dis$ and $f = g \o \piv$, so the conclusion follows.
\end{proof}

We now show that $\Phi$ respects the equivalence relation $\sim$, so that it extends to the quotient.

\begin{proposition}	\label{prop-phiprimiswelldef}
For all  $f,g \in M$, if    $f \sim g$ then 
$\Phi ( f ) \sim \Phi ( g )$. 
\end{proposition}
\vproofspace
\begin{proof}
Let $ f,g \in M$ be such that $f \sim g$. 
If $f=g$, then clearly $\Phi (f) = \Phi (g)$. 
Now suppose there is a  $\pi$  such that $f = g \o \piv$, and that $f$ (and hence also $g$) has a unique voter with maximal force. 
We need to show that $ \Phi ( f )  \sim  \Phi ( g )$. 
In fact, we will show that $\Phi ( f ) = \Phi ( g ) \o \piv$. 
From Lemma~\ref{lem-techpirel} 
we know that
$ \Inf_{\hspace{-0.1cm}\Dis}^i [ g ] =  \Inf_{\hspace{-0.1cm}\Dis}^{\pi (i)} [ f ]$ for each $i$. 
This implies that 
$\pi$ maps the first (and only) most forceful voter of $g$ 
to the first (and only) most forceful voter of $f$.
Similarly, $\pi$ maps the least forceful voters in $g$ 
to the least forceful voters in $f$.
Applying  the definition of $\Phi$ (see Definition~\ref{def-contrctphi}), 
we see that
  $\Phi ( f ) = \Phi ( g ) \o \piv$. 
\end{proof}

%\submission{}{
%\begin{proof}
%Let $ f,g \in M$ be such that $f \sim g$. 
%If $f=g$, then clearly $\Phi (f) = \Phi (g)$. 
%Now suppose there is a  $\pi$  such that $f = g \o \piv$, and that $f$ (and hence also $g$) has a unique voter with maximal force. 
%We need to show that $ \Phi ( f )  \sim  \Phi ( g )$. 
%In fact, we will show that $\Phi ( f ) = \Phi ( g ) \o \piv$. 
%From Lemma~\ref{lem-techpirel} 
%we know that
%Note that $f = g \o \piv$
%implies that 
%$ \Inf_D^i [ g ] =  \Inf_D^{\pi (i)} [ f ]$ for each $i$. 
%This implies that 
%$\pi$ maps the first (and only) most forceful voter of $g$ 
%to the first (and only) most forceful voter of $f$.
%This together with 
%the fact that $\pi$ maps the first most forceful voter of $g$ 
%to the first most forceful voter of $g$ (since by hypothesis there is only one 
%most forceful voter), 
%Similarly, $\pi$ maps the least dominant voters in $g$ 
%to the least dominant voters in $f$.
%Similarly, $\pi$ maps the least forceful voters in $g$ 
%to the least forceful voters in $f$.
%Applying  the definition of $\Phi$ (see Definition~\ref{def-contrctphi}), 
%we see that
%  $\Phi ( f ) = \Phi ( g ) \o \piv$. 
%Let $f,g \in M$ be such that $f \sim g$. 
%We need to show that $\Phi ( f ) \sim \Phi ( g )$. 
%Let $\pi$ be such that $f = g \o \pi$. 
%We will show that $\Phi ( f ) = \Phi ( g ) \o \pi$. 
%It is easy to see that $ \Inf_D^i [ g ] =  \Inf_D^{\pi (i)} [ f ]$ for each $i$.  
%\end{proof}
%}

 \begin{definition}	\label{def-contrctphiprim}
%Given a distribution $D$ on $\LA^n$, 
We  define $\Phi_\sim \colon \MExt \to \MExt$ by 
$\Phi_\sim ( [ f ] ) = [ \Phi ( f ) ]$.
%for all $[f] \in \MExt$. 
%We  define $\PhiD' \colon \MExt \to \MExt$ by 
%$\PhiD' ( [ f ] ) = [ \PhiD ( f ) ]$ 
%for all $[f] \in \MExt$. 
 \end{definition}
 
% This map is written   $\PhiExt$ if $D$ is assumed. 
 We want to point out that, 
 although we have not written it explicitly, 
 $\Phi_\sim$ (and $\Phi$ alike) does depend on 
 a chosen distribution $\Dis$. 
%Again, if $D$ is assumed, this map is also written  as $\PhiExt$.  
By Proposition~\ref{prop-phiprimiswelldef}, $\PhiExt$ is well-defined. 
%For any map $F \colon X \to X$, by $F^{(k)}$ we mean the 
%$k$-fold composition $F \o \cdots \o F$. 
%In particular, as per usual we have $F^{(0)} = 1$ and $F^{(1)} = F$. 
We also define for each $i = 1 , 2 , \ldots , n$ a map $s_i \colon \LA^n \to \LA^n$ by 
$s_i  (x_1 , \ldots , x_n ) = (x_i , \ldots , x_i )$.
% and $x_i \in \LA$. 

The following is a technical result. 

\begin{lemma}	\label{lem-techlemmaunmiv}
For any $f \in M$ and 
%$i = 1 , 2, \ldots , n$, 
$i$, voter $i$ is the unique voter with maximal force on $f \o s_i$.
%has a unique voter with maximal force. 
Moreover, 
%Further,
 $  [ f \o s_i ] =  \{ f \o s_k \mid k = 1 , 2, \ldots , n \}$ for each $i$.
\end{lemma}
\vproofspace
\begin{proof}
Since $f$ is Pareto, we have
$$ \Inf_{\hspace{-0.1cm}\Dis}^j [ f \o s_i ]
= \Prob_{x \sim \Dis} [ f (s_i (x) ) = x_j ]
=   \Prob_{x \sim \Dis} [ x_i = x_j ]  $$
for any $j$. 
Now as $\Dis$ is assumed to have full support,  $\Prob_{x \sim \Dis} [ x_i = x_j ]=1$ if{f} $x_i = x_j$ for 
all $x \in \LA^n$, i.e.,   if{f} $i=j$. 
This proves that $i$ 
is the unique voter with maximal force. 

From the first part we know that $f \o s_i$ has a unique voter with maximal force. 
Hence we have  $[ f \o s_i ] = \{  ( f \o s_i ) \o \piv \mid \pi \} = \{ f \o s_k \mid k = 1 , 2, \ldots , n \}$. 
\end{proof}

%\begin{proof}
%Since $f$ is Pareto, we have
%$$ \Inf_D^j [ f \o s_i ]
%= \Prob_{x \sim D} [ f (s_i (x) ) = x_j ]
%=   \Prob_{x \sim D} [ x_i = x_j ]  $$
%for any $j$. 
%Now as $D$ is assumed to have full support,  $\Prob_{x \sim D} [ x_i = x_j ]=1$ if{f} $x_i = x_j$ for 
%all $x \in \LA^n$, i.e.,   if{f} $i=j$. 
%This proves that $i$ 
%is the unique voter with maximal force. 
%
%From the first part we know that $f \o s_i$ has a unique voter with maximal force. 
%Hence we have  $[ f \o s_i ] = \{  ( f \o s_i ) \o \piv \mid \pi \} = \{ f \o s_k \mid k = 1 , 2, \ldots , n \}$. 
%\end{proof}

%\begin{proposition}	\label{prop-calculateqrelset}
%For any 
%$i$,
%%$i= 1 , 2, \ldots , n$,  
%$  [ f \o s_i ] =  \{ f \o s_k \mid k = 1 , 2, \ldots , n \}$.
%\end{proposition}
%\begin{proof}
%From Proposition~\ref{lem-techlemmaunmiv} 
%we know that $f \o s_i$ has a unique voter with maximal force.
%Hence, $[ f \o s_i ] = \{  ( f \o s_i ) \o \piv \mid \pi \} = \{ f \o s_k \mid k = 1 , 2, \ldots , n \}$. 
%  
%\end{proof}

\begin{proposition}	\label{prop-phihasuniquefixp}
%For every distribution $D$ on $\LA^n$  
%(with full support and  with $D = D \o \piv$ for all permutations $\pi$), 
%The map $\PhiD'$ has a unique fixpoint.
The map $\PhiExt \colon M_\sim \to M_\sim$ has a unique fixpoint.
\end{proposition}
\vproofspace
\begin{proof}
%By a corollary of Banach's fixpoint theorem (see, e.g.,~\cite{pata2014fixed})    it suffices to show that
% there exists 
%%there exists a $D$ and 
%a $k$ such that the $k$-fold composition $\PhiExt^{(k)}$ is a contraction on $\MExt$. 
%%To be precise, we show that $\PhiD'^{(n)}$ is a contraction on $\MExt$.  
%To be precise, we show that 
%$\PhiExt^{(n)}$ is a contraction on $\MExt$.  
Since $M_\sim$ is finite, 
it suffices to show  
% it holds that 
$ d_{\sim} (  \PhiExt^{(n)} ( [ f ] ) ,  \PhiExt^{(n)} ( [ g ] ) )
< d_{\sim} ( [f] , [g] )$ 
for all $[f],[g] \in M_\sim$. 
So let $[f] , [ g ] \in \MExt$. 
Every iteration of $\Phi$ on $f$, at least one voter loses their vote 
as it is taken over by the most forceful voter. 
Thus, there is an $i$ such that 
$\Phi^{(n)} ( f  ) ( x_1 , \ldots   , x_n ) 
= f ( x_i , \ldots , x_i )$ for all $(x_1 , \ldots , x_n)$,
or in other words, 
%and similarly for $g$ there is a $j$ such that 
%$\Phi^{(n)} ( g  ) ( x_1 , \ldots   , x_n ) 
%= f ( x_j , \ldots , x_j )$ for all $(x_1 , \ldots , x_n)$.
%Defining a map $s_i$ by $s_i \colon (x_1 , \ldots , x_n ) \mapsto (x_i , \ldots , x_i )$
%for each $i = 1 , 2 , \ldots , n$, 
%Thus, there is an $i$ such that 
%we can express this as saying that
$\Phi^{(n)} ( f  ) = f \o s_i$.
There is similarly a $j$ such that 
$\Phi^{(n)} ( g  ) = g \o s_j$.

We have
%\begin{eqnarray*}
%d_{\sim} (  \PhiExt^{(n)} ( [ f ] ) ,  \PhiExt^{(n)} ( [ g ] ) ) & = &  d_{\sim} (  [ \Phi^{(n)} (  f  ) ] , [ \Phi^{(n)} ( g  )  ] ) \\
%& = & d_\sim ( [ f \o s_i ] , [ g \o s_j ] ) \\
%%& & \quad  (\mbox{CLAIM 1)} \\ 
%& = & d_\sim (  \{ f \o s_k \mid k = 1 , \ldots , n \} ,  \{ g \o s_l \mid l = 1 , \ldots , n \} ) \\
%& = & \min_{k,l} d ( f \o s_k , g \o s_l ) \\
%& \leq & \min_{k} d ( f \o s_k , g \o s_k ).
%\end{eqnarray*}
%\begin{eqnarray*}
%d_{\sim} (  \PhiExt^{(n)} ( [ f ] ) ,  \PhiExt^{(n)} ( [ g ] ) ) & = &  d_{\sim} (  [ \Phi^{(n)} (  f  ) ] , [ \Phi^{(n)} ( g  )  ] ) \\
%& = & d_\sim ( [ f \o s_i ] , [ g \o s_j ] ).
%%& = & d_\sim (  \{ f \o s_k \mid k = 1 , \ldots , n \} ,  \{ g \o s_l \mid l = 1 , \ldots , n \} ).
%\end{eqnarray*}
$$
d_{\sim} (  \PhiExt^{(n)} ( [ f ] ) ,  \PhiExt^{(n)} ( [ g ] ) )  
=   d_{\sim} (  [ \Phi^{(n)} (  f  ) ] , [ \Phi^{(n)} ( g  )  ] ) 
 =  d_\sim ( [ f \o s_i ] , [ g \o s_j ] ).
$$
%Note that in the last step we made use of Lemma~\ref{lem-techlemmaunmiv}.  
%There are two cases:
%\begin{itemize}
%\item Both $f \o s_i$ as well as $g \o s_j$ does not have a unique maximal force voter. In that case, $[ f \o s_i ] = \{ f \o s_i \}$ and $[ g \o s_j ] = \{ g \o s_j \}$.
%\end{itemize}
%& & \quad  (\mbox{CLAIM 1)} \\ 
%& = & d_\sim (  \{ f \o s_k \mid k = 1 , \ldots , n \} ,  \{ g \o s_l \mid l = 1 , \ldots , n \} ).
%& = & \min_{k,l} d ( f \o s_k , g \o s_l ) \\
%& \leq & \min_{k} d ( f \o s_k , g \o s_k ).
%\end{eqnarray*}
Applying Proposition~\ref{prop-formmetricreduces} and Lemma~\ref{lem-techlemmaunmiv}, we  obtain 
$ d_{\sim} (  \PhiExt^{(n)} ( [ f ] ) ,  \PhiExt^{(n)} ( [ g ] ) ) =
 \min_{k,l} d ( f \o s_k , g \o s_l ). $
 Since $f$ and $g$ are Pareto, we have 
 $$ (f \o s_k ) ( x_1 , \ldots , x_n ) 
 = f ( x_k , \ldots , x_k )
 = x_k \qquad \mbox{ and } \qquad
  (g \o s_l ) ( x_1 , \ldots , x_n ) 
 = g ( x_l , \ldots , x_l )
 = x_l
 $$
% \begin{eqnarray*}
% & (f \o s_k ) ( x_1 , \ldots , x_n ) 
% = f ( x_k , \ldots , x_k )
% = x_k \\
% & (g \o s_l ) ( x_1 , \ldots , x_n ) 
% = g ( x_l , \ldots , x_l )
% = x_l
% \end{eqnarray*} 
 for all $(x_1 , \ldots , x_n) \in \LA^n$. 
 Thus, 
 $  \min_{k,l} d ( f \o s_k , g \o s_l )
 = \min_{k,l}    \Prob_{x \sim \Dis} [ x_k \neq x_l ] = 0, $
 and from this it follows that for all $d_{\sim} ( [f] , [g] ) \neq 0$ we have
 $ d_{\sim} (  \PhiExt^{(n)} ( [ f ] ) ,  \PhiExt^{(n)} ( [ g ] ) )
  = 0 < d_{\sim} ( [f] , [g] ). $
%  This completes the proof.
So $\Phi_\sim$ collapses all equivalence classes after $n$ steps. 
Note that $\PhiExt^{(n)} ([f]) = \DICT$ for all  $f \in M$.
%  Since $\MExt$ is a compact metric space (as it is finite),
%  the latter strict inequality 
%  $ d_{\sim} (  \PhiExt^{(n)} ( [ f ] ) ,  \PhiExt^{(n)} ( [ g ] ) ) 
%  < d_{\sim} ( [f] , [g] ) $ 
%  for all $[f], [g] \in \MExt$ 
% entails that
% the map  $  \PhiExt^{(n)}$ is a contraction. 
%   
\end{proof}

We now show that this fixpoint is the equivalence class of dictators.

%and Banach's theorem it follows that $\Phi_\sim$ has a unique fixpoint. We now show that this fixpoint is the equivalence class of dictators.

\begin{proposition}	\label{prop-dictisfixpoint}
It holds that $\Phi_\sim ( \DICT ) = \DICT$.
\end{proposition}
\vproofspace
\begin{proof}
%We show that $\Phi ( \Dict_i  ) \sim \Dict_i $ for any $i$.
%Voter $i$ is the unique most forceful voter of $\Dict_i$.
%It  is irrelevant how others vote, and 
% $\Phi ( \Dict_i  )  (x_1 , \ldots , x_n )   = x_i$ for all $(x_1, \ldots , x_n)$. Hence
%  we have  that 
% $\Phi ( \Dict_i  ) = \Dict_i$. 
From Proposition~\ref{prop-dictisfixpointerst} 
we know that $\Phi ( \Dict_i  ) \sim \Dict_i $ for each $i$.
 This implies the equality 
 $ \Phi_\sim( \DICT ) = \Phi_\sim ( [ \Dict_i ] )
 = [ \Phi ( \Dict_i ) ]
 = [ \Dict_i ] = \DICT$, 
 where we used Proposition~\ref{prop-dictformseqclass}.
\end{proof}

%\begin{proposition}	\label{prop-piaaodicttransf}
%For $n \geq 2$, 
%if $\PIIA^{n-1} \neq \DICT^{n-1}$ 
%then 
%there exists an 
%% voting rule 
%$f  \colon \LA^{n} \to \LA$
%with 
%   $f \in  \PIIA \setminus \DICT$ such that $\Phi ' ( [ f ] ) = [f]$.
%\end{proposition}
%\begin{proof}
%Let $g $ be a voting rule for $n-1$ voters, with 
% $g  \in  \PIIA^{n-1} \setminus \DICT^{n-1}$. 
%% Let 
%Define $f \colon  \LA^{n} \to \LA$ by
%$f (x_1 , \ldots , x_{n-1}, x_n)  = g( x_1 , \ldots , x_{n-1} )$. 
%Note the following. 
%\begin{enumerate}
%\item[(1)] $f \in \PIIA$ since it satisfies IIA and Pareto, as $g$ does.
%\item[(2)]  $f$ is not a dictator: clearly none of the first $n-1$ voters can be the 
%dictator for that would imply that $g$ were a dictator, and if the $n$-th voter were the dictator then 
%$ g(x_1 , \ldots , x_{n-1}) = x_n$ for all $(x_1 , \ldots , x_{n-1} , x_n)$, in contradiction
%with the fact that  
% $g$ is a function. 
%\end{enumerate}
%Thus, $f \in  \PIIA \setminus \DICT$. 
%
%%We show that $\Phi ' ( [ f ] ) = [f]$. 
%%It suffices to establish that $\Phi ( f ) = f$. 
%It is not hard to see that 
%there is a distribution $D$ such that, 
%with respect to $D$, voter $n$ is the unique voter 
%with least force on $f$.
%This shows that $\Phi ( f ) = f$. 
%Hence, $\Phi ' ( [ f ] ) = [f]$. 
%  
%\end{proof}

We shall now 
%introduce some new terminology 
%that is necessary to 
develop a technical result needed
for proving Arrow's theorem.
%Given any distribution $D$ on $\LA^{n-1}$, we can extend $D$ to 
%a distribution $D'$ on $\LA^{n}$ by 
%defining
%\begin{equation}	\label{eq-defdistrpri}
% D' ( x_1 , \ldots , x_{n-1} , x_n )
%= \frac{D(x_1, \ldots , x_{n-1})}{| \LA|}.
%\end{equation}
For any $i $, we 
define $v_i \colon \LA^n \to \LA^{n-1} $ 
by $v_i (x) = x_{-i}$, where 
$x_{-i}$ is the same vector as $x$ 
but with the $i$-th component left out 
(so $x = (x_i, x_{-i})$ for all $x$ and $i$). 
Given any distribution $\Dis$ on $\LA^{n-1}$ and 
 $i = 1 , 2, \ldots , n$,  we can 
 associate with $\Dis$ 
a real-valued map
%distribution
$\Dis^{[i]}$ on $\LA^{n}$ by 
defining
%\begin{equation*}	\label{eq-defdistrpri}
%D^{[i]} ( x ) = \frac{\sum_{\tau \colon n \to n} D ( v_i ( \tauv (x ) ) )}{n! \, | \LA |}. 
%\end{equation*}
\begin{equation}	\label{eq-defdistrpri}
\Dis^{[i]} ( x ) = \frac{\sum_{\tau \colon n \to n} \Dis ( v_i ( \tauv (x ) ) )}{n! \, | \LA |}. 
\end{equation}

%Note that 
%$\frac{\sum_{\tau \colon n \to n} ( D \o v_i \o \tauv ) (x)}{n! \, | \LA |}$.

%Now let us suppose that 
% $g \colon  \LA^{n-1} \to \LA$ is a voting rule for $n-1$ voters,
%%that is Pareto, 
%and let 
% $f \colon  \LA^{n} \to \LA$ be
%$f (x_1 , \ldots , x_{n-1}, x_n)  = g( x_1 , \ldots , x_{n-1} )$. 

% If D is (n-1)-permutation-invariant, then D' is n-permutation-invariant.

\begin{lemma}	\label{lem-laatsteleiieemeoj}
Let $\Dis$ be a distribution on $\LA^{n-1}$ and $i \in\{ 1 , 2, \ldots , n \}$.
Then $\Dis^{[i]}$ is a distribution on $\LA^{n}$ and $\Dis^{[i]}$ is $n$-permutation-invariant.
%If furthermore $\Dis$ has full support, then $\Dis^{[i]}$ has full support as well.
If moreover $\Dis$ has full support, then also $\Dis^{[i]}$ has full support.
% as well.
%If $D$ is $n-1$-permutation-invariant, then $D'$ is $n$-permutation-invariant.
\end{lemma}
\vproofspace
\begin{proof}
To show that $\Dis^{[i]}$ is a distribution, 
note that 
 for a fixed permutation $\tau \colon n \to n$, 
$$  \sum_{x \in \LA^n}  \Dis ( v_i ( \tauv (x ) ) )
=  \sum_{x \in \LA^n}  \Dis ( v_i ( x  ) )
= \sum_{x_i} \sum_{x_{-i}} \Dis (x_{-i} )
= \sum_{x_i} \, 1 
= | \LA |. $$
Thus, 
as the number of permutations $n \to n$ is $n!$, we get  
$$ \sum_{x \in \LA^n} \sum_\tau \Dis^{[i]} (x) 
= \sum_\tau  \sum_{x \in \LA^n}  \Dis ( v_i ( \tauv (x ) ) ) 
= n! \, |\LA|. $$

We now show that $\Dis^{[i]}$ is $n$-permutation-invariant. 
Let $\pi \colon n \to n$. 
We show that $\Dis^{[i]} \o \piv = \Dis^{[i]}$. 
For any $x \in \LA^n$, we have
$$ ( n! \, | \LA | ) \, ( (\Dis^{[i]} \o \piv ) ( x )  ) = 
 \sum_{\tau \colon n \to n} \Dis ( v_i ( \tauv ( \piv (x) ) ) )
=  \sum_{\sigma \colon n \to n} \Dis ( v_i ( \sigmav(x) ) ) 
= ( n! \, | \LA | ) \, ( \Dis^{[i]}  (x) ). 
$$

The last claim is trivial, so the proof is complete. 
\end{proof}

The following is a technical lemma. 

\begin{lemma}	\label{lem-laatstelemeoj}
Let $n \geq 2$, and
let $g \colon  \LA^{n-1} \to \LA$ be a voting rule for $n-1$ voters. 
%that is Pareto, 
If $f \colon  \LA^{n} \to \LA$ is such that
$f (x_1 , \ldots , x_{n-1}, x_n)  = g( x_1 , \ldots , x_{n-1} )$, 
and $\Dis$ is a $n-1$-permutation-invariant distribution on $\LA^{n-1}$,
% and let $D' = D^{[n]}$.
% be as in the above. 
then %we have
$\Inf_{\hspace{-0.1cm}\Dis^{[n]}}^n [f]  \leq 2/ ( n \, | \LA |)$, 
and 
 for each $i = 1 , 2 , \ldots , n-1$ that
$\Inf_{\hspace{-0.1cm}\Dis^{[n]}}^i [f] \geq \frac 1 n   \Inf_{\hspace{-0.1cm}\Dis}^i [ g ]$.
%and $\Inf_{\hspace{-0.1cm}\Dis^{[n]}}^n [f]  \leq 1/ | \LA |$.
\end{lemma}
\vproofspace
\begin{proof}
Let $i = 1 , 2 , \ldots , n-1$. 
We show that
$\Inf_{\hspace{-0.1cm}\Dis^{[n]}}^i [f] \geq \frac 1n  \Inf_{\hspace{-0.1cm}\Dis}^i [ g ]$. 
More precisely, we will show that 
$$ \Inf_{\hspace{-0.1cm}\Dis^{[n]}}^i [f]  = 
 \frac 1n  \Inf_{\hspace{-0.1cm}\Dis}^i [ g ] + 
  \frac{1}{n \, | \LA | } \sum_{j=1}^{n-1}  \mu ( x_{-j} ) \, 1_{g(x_1  , \ldots , x_{n-1} ) = x_i}. $$
We have 
$$ \Inf_{\hspace{-0.1cm}\Dis^{[n]}}^i [f] 
= \frac{1}{n! \, | \LA | } \sum_{x_1 , \ldots , x_n} \sum_{\tau} 
\mu ( x_{\tau (1)} , \ldots , x_{\tau (n-1)} ) \, 1_{g(x_1  , \ldots , x_{n-1} ) = x_i}.
$$
%Let $V =  \{ \tau \colon n \to n \mid \tau ( n ) = n \}$.
%We have
%\begin{eqnarray*}
%& & \sum_{x_1 , \ldots , x_{n-1} , x_n } \, \sum_{\tau \in V}   \mu ( x_{\tau (1)} , \ldots , x_{\tau (n-1)} ) \, 1_{g(x_1  , \ldots , x_{n-1} ) = x_i}  \\
%& = &  \sum_{x_1 , \ldots , x_{n-1} , x_n } \, \sum_{\tau \in V}   \mu ( x_1 , \ldots , x_{n-1} ) \, 1_{g(x_1  , \ldots , x_{n-1} ) = x_i} \\
%& = & (n-1)! \,   \sum_{x_1 , \ldots , x_{n-1} , x_n }   \mu ( x_1 , \ldots , x_{n-1} ) \, 1_{g(x_1  , \ldots , x_{n-1} ) = x_i} \\
%& = & (n-1)! \,   \sum_{x_1 , \ldots , x_{n-1} } \sum_{x_n}   \mu ( x_1 , \ldots , x_{n-1} ) \, 1_{g(x_1  , \ldots , x_{n-1} ) = x_i} \\
%& = &  (n-1)! \, | \LA |  \sum_{x_1 , \ldots , x_{n-1} }  \mu ( x_1 , \ldots , x_{n-1} ) \, 1_{g(x_1  , \ldots , x_{n-1} ) = x_i} \\
%& = &  (n-1)! \, | \LA |   \Inf_{\hspace{-0.1cm}\Dis}^i [ g ]
%\end{eqnarray*} 
%Since
%$$ \Inf_{\hspace{-0.1cm}\Dis^{[n]}}^i [f] \geq
% \frac{1}{n! \, | \LA | } \sum_{x_1 , \ldots , x_n} \sum_{\tau \in V} 
%\mu ( x_{\tau (1)} , \ldots , x_{\tau (n-1)} ) \, 1_{g(x_1  , \ldots , x_{n-1} ) = x_i},$$
%the bound follows. 
 For each $i = 1 , 2 , \ldots , n$, let $V_i = \{ \tau \colon n \to n \mid \tau ( i ) = n \}$. 
To start, note that 
\begin{eqnarray}
& & \sum_{x_1 , \ldots , x_{n-1} , x_n } \, \sum_{\tau \in V_n}   \mu ( x_{\tau (1)} , \ldots , x_{\tau (n-1)} ) \, 1_{g(x_1  , \ldots , x_{n-1} ) = x_i}  \nonumber \\
\text{\small (since $\Dis$ is $n-1$-perm.inv.)}& = &  \sum_{x_1 , \ldots , x_{n-1} , x_n } \, \sum_{\tau \in V_n}   \mu ( x_1 , \ldots , x_{n-1} ) \, 1_{g(x_1  , \ldots , x_{n-1} ) = x_i}  \nonumber \\
\text{\small (since $|V_n| = (n-1)!$)} & = & (n-1)! \,   \sum_{x_1 , \ldots , x_{n- 1} , x_n }   \mu ( x_1 , \ldots , x_{n-1} ) \, 1_{g(x_1  , \ldots , x_{n-1} ) = x_i}   \nonumber \\
& = & (n-1)! \,   \sum_{x_1 , \ldots , x_{n-1} } \sum_{x_n}   \mu ( x_1 , \ldots , x_{n-1} ) \, 1_{g(x_1  , \ldots , x_{n-1} ) = x_i}   \label{letsteqo}  \\
& = &  (n-1)! \, | \LA |  \sum_{x_1 , \ldots , x_{n-1} }  \mu ( x_1 , \ldots , x_{n-1} ) \, 1_{g(x_1  , \ldots , x_{n-1} ) = x_i}  \nonumber \\
& = &  (n-1)! \, | \LA |   \Inf_{\hspace{-0.1cm}\Dis}^i [ g ].  \nonumber
\end{eqnarray} 
 Also, for each $i \neq n$, note that 
 $V_i = \bigcup_{j=1}^{n-1} H_j^i$ where
 $$ H_j^i = \{ \tau \mid  \tau (i) = n \mbox{ and } \tau (n) = j \}. $$
 It is clear that $|  H_j^i  | = (n-2)!$.
 If $\tau \in H_j^i$, then
 clearly  $\mu ( x_{\tau (1)} , \ldots , x_{\tau(n-1)} ) 
 = \mu ( x_{-j})$. 
Now fix $x_1 , \ldots , x_{n-1} , x_n$, and a $j \in \{ 1 , 2 , \ldots , n-1\}$. 
Then 
\begin{eqnarray*}
\sum_{i=1}^{n-1} \sum_{\tau \in H_j^i}  \mu ( x_{\tau (1)} , \ldots , x_{\tau (n-1)} ) \, 1_{g(x_1  , \ldots , x_{n-1} ) = x_i} 
& = & \sum_{i=1}^{n-1} \sum_{\tau \in H_j^i}  \mu ( x_{-j}) \, 1_{g(x_1  , \ldots , x_{n-1} ) = x_i} \\
& = & (n-1) | H_j^i |  \mu ( x_{-j}) 1_{g(x_1  , \ldots , x_{n-1} ) = x_i} \\
& = & (n-1) (n-2)! \mu ( x_{-j}) 1_{g(x_1  , \ldots , x_{n-1} ) = x_i} \\
& = & (n-1)! \mu ( x_{-j}) 1_{g(x_1  , \ldots , x_{n-1} ) = x_i}.
\end{eqnarray*}
Summing this expression over all $j = 1 , 2 , \ldots , n-1$, we get
$$  (n-1)! \,  1_{g(x_1  , \ldots , x_{n-1} ) = x_i}   \sum_{j=1}^{n-1} \mu ( x_{-j}).$$
Thus,
\begin{eqnarray*}
 \Inf_{\hspace{-0.1cm}\Dis^{[n]}}^i [f]  & =& \frac{1}{n! \, | \LA | } \sum_{x_1 , \ldots , x_n} \sum_{\tau} 
\mu ( x_{\tau (1)} , \ldots , x_{\tau (n-1)} ) \, 1_{g(x_1  , \ldots , x_{n-1} ) = x_i} \\
& = &  \frac{1}{n! \, | \LA | } \sum_{x_1 , \ldots , x_n} \sum_{\tau \in V_n} 
\mu ( x_{\tau (1)} , \ldots , x_{\tau (n-1)} ) \, 1_{g(x_1  , \ldots , x_{n-1} ) = x_i} \\ 
& & + \,   \frac{1}{n! \, | \LA | } \sum_{x_1 , \ldots , x_n} \sum_{\tau \in \cup_{i=1}^{n-1} V_i} 
\mu ( x_{\tau (1)} , \ldots , x_{\tau (n-1)} ) \, 1_{g(x_1  , \ldots , x_{n-1} ) = x_i}.
%& = &  \frac 1n  \Inf_{\hspace{-0.1cm}\Dis}^i [ g ] +  \frac{1}{n! \, | \LA | } \sum_{x_1 , \ldots , x_n} \sum_{j=1}^{n-1} \sum_{i=1}^{n-1} \sum_{\tau \in H_j^i}  \mu ( x_{\tau (1)} , \ldots , x_{\tau (n-1)} ) \, 1_{g(x_1  , \ldots , x_{n-1} ) = x_i}.
\end{eqnarray*}
Plugging in our calculations from above, we conclude that 
$$ \Inf_{\hspace{-0.1cm}\Dis^{[n]}}^i [f]  = 
 \frac 1n  \Inf_{\hspace{-0.1cm}\Dis}^i [ g ] + 
  \frac{1}{n \, | \LA | } \sum_{j=1}^{n-1}  \mu ( x_{-j} ) \, 1_{g(x_1  , \ldots , x_{n-1} ) = x_i}. $$
%$$ 
%\Inf_{\hspace{-0.1cm}\Dis^{[n]}}^i [f] 
%\geq  \frac{1}{n! \, | \LA |} ( (n-1)! | \LA  | \sum_{x_1 , } $$ 

If we repeat the steps from above, but with $i$ replaced by $n$, then the whole reasoning is the same, except in~\eqref{letsteqo}: there, we get 
$$  (n-1)! \,   \sum_{x_1 , \ldots , x_{n-1} } \sum_{x_n}   \mu ( x_1 , \ldots , x_{n-1} ) \, 1_{g(x_1  , \ldots , x_{n-1} ) = x_n} 
=  (n-1)! \,   \sum_{x_1 , \ldots , x_{n-1} } \mu ( x_1 , \ldots , x_{n-1} ) = (n-1)!.
$$ 
This lets us conclude that 
$$ \Inf_{\hspace{-0.1cm}\Dis^{[n]}}^n [f]  = 
  \frac{1}{n \, | \LA | }  + 
  \frac{1}{n \, | \LA | } \sum_{j=1}^{n-1}  \mu ( x_{-j} ) \, 1_{g(x_1  , \ldots , x_{n-1} ) = x_n}. $$
Since $ \sum_{j=1}^{n-1}  \mu ( x_{-j} ) \, 1_{g(x_1  , \ldots , x_{n-1} ) = x_n} \leq 1$ we have
$ \Inf_{\hspace{-0.1cm}\Dis^{[n]}}^n [f]  \leq 2/ ( n \, | \LA |)$. 
\end{proof}

Let $\varepsilon>0$ be 
%a small number 
%(below we shall specifically need 
such that 
$\varepsilon < 1 - 2/| \LA|$. 
Note that this is possible precisely because $| \LA | > 2$; 
this is where we use the assumption that $|A| \geq 3$, i.e., 
there are at least three candidates. 
Fix any $y \in \LA$. 
Let $G = \{ (y, \ldots , y ) \}$.
% \mid x \in \LA \} \subseteq \LA^{n-1}$. 
We define a particular distribution $\Dis_*$ on $\LA^{n-1}$, as follows: 
$\Dis_*$ gives  weight $1-\varepsilon$
to $(y, \ldots , y )$ and spreads the remaining 
$\varepsilon$ out over all other profiles. 
That is, for any $x \in \LA^{n-1}$, 
we
let 
$$\Dis_* (x )  =
\begin{cases} 1-\varepsilon & \mbox{ if $x \in G$}, \\
 \frac{\varepsilon}{|\LA|^{n-1}-1} & \mbox{ if $x \not\in G$}.
\end{cases}
$$
Note that $\Dis_*$ is $n-1$-permutation-invariant.  
%Note that for Pareto $g \colon \LA^{n-1} \to \LA$, $ g(x, x, \ldots , x)=x$ for all $x \in \LA$. 
%Note that for Pareto $g \colon \LA^{n-1} \to \LA$, 
Also note that for a Pareto $g \colon \LA^{n-1} \to \LA$ we have 
$g (x, x , \ldots , x ) = x $ for all $x \in \LA$.

\begin{lemma}	\label{lem-laatstelem}
Let $g \colon  \LA^{n-1} \to \LA$ be a voting rule for $n-1$ voters 
that is Pareto, and let 
 $f \colon  \LA^{n} \to \LA$ be
$f (x_1 , \ldots , x_{n-1}, x_n)  = g( x_1 , \ldots , x_{n-1} )$. 
Then it holds that voter $n$ is the unique least forceful voter of $f$ 
with respect to ${\Dis_*^{[n]}}$.
%(where ${D_*^{[n]}}$
%is the distribution on $\LA^n$ obtained from $D_*$ 
%as explained in the above).
\end{lemma}
\vproofspace
\begin{proof}
By Lemma~\ref{lem-laatstelemeoj} it suffices to show 
that $ \Inf_{\hspace{-0.1cm}\Dis_*}^i [ g ] > 2/ | \LA |$ 
for each $i = 1 , 2 , \ldots , n-1$.

We calculate:
$$
\Inf_{\hspace{-0.1cm}\Dis_*}^i [ g ]  =   \sum_{(x_1, \ldots , x_{n-1})} 
\Dis_* (x_1 , \ldots , x_{n-1}) \,  1_{g(x_1 , \ldots , x_{n-1} ) = x_i} 
 \geq  \sum_{(x_1, \ldots , x_{n-1}) \in G} \Dis_* (x_1 , \ldots , x_{n-1}) \,  1_{g(x_1 , \ldots , x_{n-1} ) = x_i},
 $$
% \end{eqnarray*}
% \begin{eqnarray*}
%\Inf_{D_*}^i [ g ] & = &  \sum_{(x_1, \ldots , x_{n-1})} 
%D(x_1 , \ldots , x_{n-1}) \,  1_{g(x_1 , \ldots , x_{n-1} ) = x_i} \\
% & \geq & \sum_{(x_1, \ldots , x_{n-1}) \in G} D(x_1 , \ldots , x_{n-1}) \,  1_{g(x_1 , \ldots , x_{n-1} ) = x_i},
% \end{eqnarray*}
and this equals
$ \Dis_* (y, \ldots , y ) = 1 - \varepsilon$ 
since $g$ is Pareto. 
By choice of $\varepsilon$, we have $1- \varepsilon > 2/ | \LA |$. 
\end{proof}

Finally we arrive at the proof of Arrow's theorem. 

\vspace{0.2cm}

%\begin{thom}[Arrow]	\label{thm-arrow}
%A voting rule for at least three candidates that satisfies 
%the Pareto property and IIA must be a dictatorship. 
%\end{thom}
\noindent \emph{Proof of Arrow's theorem}.
Assume towards a contradiction that 
$n \geq 2$ and there is a
$g \in \PIIA^{n-1} \setminus \DICT^{n-1}$. 
%We start by showing that 
%for any $n \geq 2$, 
%if it is the case that $\PIIA^{n-1} \neq \DICT^{n-1}$ 
%then 
%there exists an 
%% voting rule 
%$f  \colon \LA^{n} \to \LA$
%with 
%   $f \in  \PIIA \setminus \DICT$ and $\Phi ' ( [ f ] ) = [f]$. 
%If    $\PIIA^{n-1} \neq \DICT^{n-1}$, then it must 
%hold that $\PIIA^{n-1} \not\subseteq \DICT^{n-1}$, as 
%any dictator for sure is Pareto and satisfies IIA.
%So let $g $ be a voting rule for $n-1$ voters, with 
% $g  \in  \PIIA^{n-1} \setminus \DICT^{n-1}$. 
%% Let 
We define $f \colon  \LA^{n} \to \LA$ by
$f (x_1 , \ldots , x_{n-1}, x_n)  = g( x_1 , \ldots , x_{n-1} )$. 
Note the following. 
\begin{enumerate}
\item[(1)] $f \in \PIIA$ since it satisfies IIA and Pareto, as $g$ does.
\item[(2)]  $f$ is not a dictator: clearly none of the first $n-1$ voters can be the 
dictator for that would imply that $g$ were a dictator, and if the $n$-th voter were the dictator then 
$ g(x_1 , \ldots , x_{n-1}) = x_n$ for all $(x_1 , \ldots , x_{n-1} , x_n)$, in contradiction
with the fact that  
 $g$ is a function. 
\end{enumerate}
Thus, $f \in  \PIIA \setminus \DICT$.

We take $\Dis$ to be the $\Dis_*^{[n]}$ that we just introduced. 
%
%$D_{[n]}^*$
By construction,  $\Dis_*^{[n]}$ has full support, satisfies 
$\Dis_*^{[n]} = \Dis_*^{[n]} \o \piv$ for all permutations $\pi$ (by Lemma~\ref{lem-laatsteleiieemeoj}), and  voter $n$ is the unique voter 
with least force on $f$ with respect to $\Dis_*^{[n]}$ (by Lemma~\ref{lem-laatstelem}).
%, by Lemma~\ref{lem-laatstelem}.
This proves  that 
%for $\Phi_\sim^{D_* \hspace{-0.1cm}'} \colon \MExt \to \MExt$ 
%it holds that  
$ \Phi^{\Dis_*^{[n]}} ( f ) = f $.
Therefore, 
%we have
$ \Phi_\sim^{\Dis_*^{[n]}} ( [f ] ) = [f ]$. 
Proposition~\ref{prop-phihasuniquefixp}  and  Proposition~\ref{prop-dictisfixpoint}
 imply that  $[f] =\DICT$.
In particular, $f \in  \DICT$. 
Contradiction. \hfill $\square$

%%We take D to be  ?? Note that D has full support and satisfies ?.
%%Now, consider the resulting metric space M^D = (M,d_D) and the map
%%Now \Phi^D:M^D \to M^D, and its quotient M^D_\sim = (M_\sim,d_\sim) and 
%%the map \Phi^D_\sim: M^D_\sim \to M^D_\sim wrt the equiv.rel given in 
%%[Show that n is the unique least voter ... and why this entails 
%%\Phi^D(f)=f.]
%
%
%%We show that $\Phi ' ( [ f ] ) = [f]$. 
%%It suffices to establish that $\Phi ( f ) = f$. 
%It is not hard to see that 
%there is a distribution $D$ such that, 
%with respect to $D$, voter $n$ is the unique voter 
%with least force on $f$.
%This shows that $\Phi ( f ) = f$, where $\Phi = \PhiD$. 
%Hence, $\Phi ' ( [ f ] ) = [f]$. 
% 
%   
%
%
%
%Now to finish the proof, we argue by contradiction. 
%If there were 
%an $n \geq 2$ with $\PIIA^{n-1} \neq \DICT^{n-1}$, then
%% from the above we get 
%%an $ f \in  \PIIA \setminus \DICT$ such that $\Phi ' ( [ f ] ) = [f]$
% Proposition~\ref{prop-phihasuniquefixp}  and  Proposition~\ref{prop-dictisfixpoint}
%together would imply that  $[f] =\DICT$.
%In particular, $f \in  \DICT$. 
%Contradiction.

%From   Proposition~\ref{prop-piaaodicttransf} 
%we then get an $ f \in  \PIIA \setminus \DICT$ such that $\Phi ' ( [ f ] ) = [f]$. 
%Proposition~\ref{prop-phihasuniquefixp}  and  Proposition~\ref{prop-dictisfixpoint}
%together imply that  $[f] =\DICT$.
%In particular, $f \in  \DICT$. 
%Contradiction.
%  
%\end{proof}

\section{Conclusion}		\label{sec-conclusion}

%HELLE COMMENTS on conclusion:
%The conclusion is currently not much more than a repetition of parts 
%of the introduction. In the conclusion, you should *reflect on your 
%work*, e.g. by discussing its limitations/generality, your assumptions, 
%etc.

The main goal of this paper has been to show that Arrow's impossibility theorem 
can be proved using Banach's fixpoint theorem. 
Our approach involved coming up with an appropriate equivalence relation, and then 
 defining a contraction on the resulting equivalence space whose 
unique fixpoint is the set of dictators. 
%This led to  Proposition~\ref{prop-phihasuniquefixp}, 
%%which can be interpreted as saying that 
%with the interpretation that 
%iterating a certain process on a sufficiently 
%nice voting rule leads to a dictatorship. 
The concept of force of a voter, 
as well as thinking about voting rules 
as elements of a metric space based on a probability distribution, 
are inspired by the Boolean analysis approach 
to social choice, a line of research initiated by~\cite{kalai2002fourier} 
and further developed by others~\cite{mossel2012quantitativeor}. 
% The proof we give  

Our proof of Arrow's theorem is different in spirit from 
most of the previous proofs in that it 
 does not involve 
 manipulations of specific profiles. 
% In that sense we would like to think that 
% our proof 
The Pareto property is fundamental in our analysis 
and we used it ubiquitously, as our original metric space 
consists of all Pareto voting rules. 
 Interestingly, however,  in our proof we did not use the
 IIA property explicitly,
 we only used it by noting 
 that it was preserved under a certain operation. 
 This makes us wonder if it would be 
 possible to get other impossibility results 
 by considering other properties 
 (that preserve the same operation).

%  IIA property
% is preserved. 
% when we showed it is preserved.
% that it is preserved in a certain case. 
% This suggests that our approach 
% our only use of the IIA property 
% in the proof of Arrow's theorem 
% Conceptually,  our approach advances the following novel perspective on Arrow's impossibility result: 
%Desirable election mechanisms necessarily converge towards a dictatorship, which can be 
%thought of as a fixpoint. 

An advantage of 
%the fixpoint 
our perspective on Arrow's theorem is that it 
establishes a link between this 
pivotal result of mathematical economics 
and  a concept often surfacing in that area: fixpoints. 
The notion of fixpoint also connects the theorem better to the area of computer science, 
where fixpoints are omnipresent 
and sometimes can lead to algorithms, 
although this does not seem to be the case here. 
%and can lead to algorithms.   
A possible direction for future work is to analyze if similar results in the area, such as 
the Gibbard-Satterthwaite theorem~\cite{gibbard1973manipulation},  
can be proved via a related fixpoint argument. 
It would also be worthwhile to study the relationship between our 
notion of power and  notions like 
decisiveness or pivotalness that 
other  proofs use.

 \paragraph{Acknowledgments.} 
We would like to thank Thomas Santoli 
% for  stimulating discussion, and in particular 
 for pointing out in a discussion that the permutation structure 
 gives rise to a group action, which ultimately led 
 to Proposition~\ref{prop-formmetricreduces}, and 
 Ronald de Wolf for inspiring us to 
 formalize and study the notion of force.

\bibliographystyle{eptcs}
\bibliography{socchoice-refsprim}

\end{document}